\documentclass[
 reprint,
 preprintnumbers,
 amsmath,
 amssymb,
 aps,
 prstab,
 floatfix,
]{revtex4-1}

\usepackage{graphicx}			
\usepackage{verbatim}			
\usepackage{wasysym}			
\usepackage{mathtools}			
\usepackage[makeroom]{cancel}		
\usepackage{dcolumn}			
\usepackage{hyperref}			
\usepackage{color}			

\newcommand{\pd}{\partial}				
\newcommand{\dd}{\mathrm{d}}				

\newcommand{\Nb}{N_{\text{p}}}		
\newcommand{\Qb}{ Q_{\beta}}		
\newcommand{\Qs}{ Q_s}		

\newcommand{\Qsc}{\Delta Q_\mathrm{sc}}

\begin{document}

\title{Convective Instabilities of Bunched Beams with Space Charge}
\author{A.~Burov}
\email{burov@fnal.gov}
\affiliation{Fermilab, PO Box 500, Batavia, IL 60510-5011}
\date{\today}

\begin{abstract}

For a single hadron bunch in a circular accelerator at zero chromaticity, without multi-turn wakes and without electron clouds and other beams, only one transverse collective instability is possible, the mode-coupling instability, or TMCI. For sufficiently strong space charge (SC), the instability threshold of the wake-driven coherent tune shift normally increases linearly with the SC tune shift, as independently concluded by several authors using different methods. This stability condition has, however, a very strange feature: at strong SC, it is totally insensitive to the number of particles. Thus, were it correct, such a beam with sufficiently strong SC, being stable at some intensity, would remain stable at higher intensity, regardless of how much higher! This paper suggests a resolution of this conundrum: while SC suppresses TMCI, it introduces head-to-tail convective amplifications, which could make the beam even less stable than without SC, even if all the coherent tunes are real, i.e. all the modes are stable in the conventional {\it absolute} meaning of the word. This is done using an effective new method of analysis of the beam's transverse spectrum for arbitrary space charge and wake fields. Two new types of beam instabilities are introduced: the {\it saturating convective instability}, SCI, and the {\it absolute-convective instability}, ACI.  

\end{abstract}


\maketitle

 


\section{\label{sec:ACI}Absolute and Convective Instabilities}

Transverse mode coupling instabilities (TMCI) are believed to be one of the main limitations for the intensity of bunched beams. Such single-bunch instabilities develop when the head-tail phase is small, $\xi \sigma_\delta/\Qs \ll 1$, where $\xi$ is the chromaticity, $\sigma_\delta$ is the relative rms momentum spread and $\Qs$ is the synchrotron tune, so that the chromatic effects can be neglected, see e.g. Ref.~\cite{chao1993physics}. For proton beams, an important question is: how does TMCI depend on the space charge (SC) tune shift $\Qsc$? Since the latter is typically high for low and medium energy machines, where the SC parameter $q=\Qsc/\Qs \gg 1$ even far from the transition energy, the question is really important.  

According to Ref.~\cite{blaskiewicz1998fast, burov2009head, PhysRevSTAB.14.094401, balbekov2017transverse, PhysRevAccelBeams.20.114401, PhysRevAccelBeams.21.104201}, TMCI intensity threshold increases, with rare exceptions, proportionally to the SC tune shift; SC makes the bunch more stable in this respect. We called this dominating class of the mode-coupling instabilities {\it vanishing}, meaning that they vanish at high SC, when the SC tune shift sufficiently exceeds the wake-related coherent tune shifts. For coasting beams, however, dependence of the transverse instability on SC is opposite: the threshold wake amplitude drops down as SC increases, since the latter suppresses Landau damping~\cite{PhysRevSTAB.12.034201}. This oppositeness may be especially puzzling in light of the similarity, if not identity, of the bunched and coasting beam thresholds without SC, clearly seen when the former is expressed in terms of the maximal line density and rms momentum spread, and the latter is supposed to be Gaussian or alike. This similarity was presumed long ago by approximate fast microwave transverse stability criterion for bunched beams \cite{Ruth:1981xt}. The main idea behind it (as well as behind similar Keil-Schnell-Boussard criterion for the longitudinal direction) was one of a fast microwave instability, which occurs locally, so that neither synchrotron motion nor bunch density variation should be very important. Following this idea, it can be expected that the microwave criterion is rather accurate for short wakes, and may be a reasonable estimation with correct scaling for arbitrary wakes. Indeed, the agreement of this criterion with the computed TMCI thresholds was found to be especially good for short wakes, i.e. long bunches; for details see Ref.~\cite{Metral:2004vi, Burov:2018iid}. Due to this similarity of the coasting beam stability criteria to the bunched one without SC, the theoretical statement of oppositeness of their dependence on SC raises a suspicion that something is lost in the picture. The suspicion is strengthened with realization that at strong SC the mentioned TMCI stability condition does not depend on the number of particles, since wake tune shift and space charge tune shift are equally proportional to that. Thus, if the beam is stable at some number of particles, it should remain stable at any number of them according to this criterion. This strange statement is certainly refuted by measurements at CERN SPS \cite{Bartosik:2013qji}, especially for the old Q26 optics with its high SC parameter, $q \simeq 20$, showing definite existence of the threshold number of particles and only weak, if any, dependence of the instability thresholds on SC. Non-existence of the threshold number of particles for the vanishing TMCI cases shows that, most likely, the real instability mechanism is lost in that picture. If so, what could it be?

For both, coasting and bunched beams, the microwave instability can be considered in terms of time evolution of an initial wave packet, traveling opposite to the beam motion, due to the wake causality. If the wave packet grows, as it travels that way along the coasting beam, the beam is unstable. However, this may not be so for the bunched beam, where a similarly growing microwave packet may travel in that manner only for a finite length, up to the bunch tail, and for a finite time, shorter than a half of the synchrotron period; thus, its growth does not necessarily entail the collective instability. The same mechanism of interaction may cause instability in the coasting beam and only head-to-tail signal amplification in the bunch, without making the entire bunch unstable. To articulate this distinction, we may, following Ref.~\cite{LifPitKin} and references therein, use the terms {\it absolute} and {\it convective} to qualify instabilities: for the former, the initial perturbation causes an unrestricted exponential growth everywhere in the medium, while for the latter, there is only a spacial amplification, and the perturbation eventually decays everywhere when a dissipation is added, no matter how tiny. Thus, the absolute instability in the coasting beam may correspond to only a finite amplification along the similar bunch, i.e. to a convective instability, without the absolute growth of the initial perturbation. It is worth noting that even without SC the TMCI threshold may be much higher than the instability threshold of the corresponding coasting beam: for instance, this is the situation with the air-bag bunch in a square well, the ABS \cite{blaskiewicz1998fast}. Due to the absence of Landau damping, the corresponding two-stream coasting beam instability is threshold-less, while the TMCI threshold for the ABS model is finite. 

With SC, for the dominant {\it vanishing} case, the TMCI threshold tends to grow proportionally to the SC tune shift, while the condition for the wave packet {\it to grow fast} benefits from SC due to its tendency to make the bunch slices rigid (see Ref.~\cite{burov2009head} and multiple plots below). Thus, there should be an interval of wake amplitudes between the convective and absolute instability, and the width of this interval has to increase linearly with SC. Note that contrary to the beam breakup in linacs (BBU), the expected regime of the signal amplification would not lead to an unlimited growth of the perturbation at the bunch tail; the amplification saturates due to the synchrotron oscillations. Indeed, with all the collective modes being stable, it is impossible to get an unlimited signal. However, on time intervals short compared to the synchrotron period, the amplification should grow similarly to the beam breakup. To keep this distinction, I call the convective instabilities of a bunch with its spectrum of absolutely-stable modes {\it saturating convective instabilities}, SCI, while the beam breakup represents its alternative, an {\it unbounded convective instability}, UCI, typical for bunches without modes. 

One more important feature of the convective instabilities is that they make the bunch prone to the absolute instability: even a weak tail-to-head action by means of a multibunch or over-revolution wake, negligible by itself, may be sufficient to make bunch oscillations unstable absolutely. Moreover, it is shown below that a damper of any sort, including the conventional bunch-by-bunch resistive kind, works as a generator of the absolute instability if the convective instability is present; thus, sufficiently large convective amplification turns the bunch into a sort of fragile metastable state. The absolute instability of such a combined kind may be called an {\it absolute-convective instability}, or ACI. One more possibility for the tail-to-head action relates to the bunch halo, which may lead to a special sort of ACI, the core-halo instability, see Ref.~\cite{Burov:2018rmx}. 

Amplification of the fast transverse microwave perturbation along the bunch was considered by D.~Brandt and J.~Gareyete~\cite{Brandt:188910, Gareyte:217745} with respect to a long positron bunch in the CERN SPS in LEP era; a BBU-type estimation for the amplification coefficient was found. This no-SC formula was suggested later for the proton bunch in the same machine by R.~Cappi, E.~Metral and G.~Metral~\cite{Cappi:2000ze}; based on that, the bunch lifetime with respect to the tail particle losses was estimated. It was pointed out in Ref.~\cite{Metral:2004vi}, that substitution of the inverse synchrotron frequency for this BBU lifetime leads to the same formula as the TMCI threshold, within a factor smaller than 2. On this ground, it was concluded that the same instability shows itself as TMCI, when approached from below the threshold, or as BBU, when approached from above it. The same statement has been made by J.~Gareyte~\cite{Gareyte:477074} a bit earlier. As it will be shown in this paper, that BBU-TMCI identification and distinction of Gareyte et al., being reasonable at no-SC, is incorrect when SC is strong, as it is at PS and similar low and medium energy rings. It will be also explained why the no-SC formula for TMCI threshold worked fairly well for the SPS, notwithstanding that the actual TMCI threshold was much above the number predicted by this formula.       

It is worth noting that saturating convective instabilities in the longitudinal plane are discussed for bunches with significant SC since long ago, see Ref.~\cite{PhysRevSTAB.6.034207} and references therein. Although they are not intended to be considered in this paper, a simple explanation of their existence well deserves to be presented here. Quoting Ref.~\cite{PhysRevSTAB.6.034207}, "due to the fact that the backward running (slow) mode grows and the forward running (fast) mode is damped the total growth over one round trip vanishes." In fact, similar explanation works for the transverse plane below TMCI threshold, when the bunch slices can be treated as rigid and the synchrotron motion slow compared with the wake-related phase and group velocities of microwave packets: the same logic leads in this case to the conclusion about the saturating convective instabilities in the transverse planes.   

Convective transverse instabilities of bunched beams with SC are considered in this paper by means of the ABS model. At the last two sections, it is shown how the suggested understanding, supported by various computations, resolves the mentioned paradoxes and controversies between the theory and observations of TMCI with SC.  


\section{\label{sec:ABS}ABS Model}

\subsection{Description}

So far, the ABS model is an only one which allows effective and precise solutions of Vlasov equations for transverse oscillations of bunched beams with arbitrary wake functions and SC tune shifts. Moreover, this model is realistic: the ABS bunch can be physically prepared, and many of its core features are not so different from the Gaussian bunch. For the no-SC case and broadband impedance, it yields the instability threshold fairly close to the Gaussian bunch, albeit modes of different numbers may turn out to couple first with it~\cite{Burov:2018iid}. For strong SC, spectral properties of ABS with various wakes were recently discussed in Ref.~\cite{PhysRevAccelBeams.21.104201}; generally, they were found to be not so different from those of the Gaussian bunches in parabolic potential wells. Certainly, ABS has its limitations, as any physical model. Some of them are rather obvious, like its missing of the {\it intrinsic Landau damping}~\cite{burov2009head, PhysRevSTAB.18.074401}, while others may show themselves only at later stages; some of the limitations can be effectively overcome with reasonable model modifications as in Refs.~\cite{PhysRevSTAB.6.014203, Burov:2018rmx}, others, really not. Since the main virtue of the model, a combination of its exact and effective solvability, physical reason and richness, is extraordinary, let us go ahead with it.   

It is convenient in this case to measure coordinates $s$ along the bunch as fractions of its full length, so that $0 \leq s \leq 1$, and to measure time $\theta$ in synchrotron radians, so the synchrotron period $T_s=2\pi$. The wake functions can be dimensionlessed by measuring them in units of their amplitudes, specified for each case. After that, equations of motion of the positive and negative fluxes of the ABS bunch in terms of their complex amplitudes $x^\pm(\theta, s)$ can be presented as follows 
\begin{equation}
\begin{split}
	&\frac{\pd\,x^+}{\pd \theta} -\frac{1}{\pi} \frac{\pd\,x^+}{\pd s} = \frac{i q}{2}(x^+ - x^-) + i\,F , \\	
	&\frac{\pd\,x^-}{\pd \theta} +\frac{1}{\pi} \frac{\pd\,x^-}{\pd s} = \frac{i q}{2}(x^- - x^+) +i\, F, 	\\
	& F(\theta, s) = w \int_0^s \dd s' W(s-s') \bar{x}(\theta, s') \,,
\end{split}
\label{ABS main}
\end{equation}
where the local centroid offset $ \bar{x}=(x^++x^-)/2$, and the boundary conditions 
\begin{equation}
\label{bc}
x^+ = x^- \;\; {\mathrm {at}}\; s=0,1 \,.
\end{equation}
%
Here the SC parameter $q$ is the ratio of the SC tune shift to the synchrotron tune, and $w$ is the wake parameter: 
\begin{equation}
\label{chi}
	w = \frac{\Nb W_0 r_0 R_0}{4\,\pi\,\gamma\,\beta^2\,\Qb \Qs}.
\end{equation}
with $\Nb$ as the number of particles per bunch, $W_0$ as the wake amplitude, $r_0$ as the classical radius, $R_0$ as the average radius of the machine, $\gamma$ and $\beta$ as the relativistic factors, $\Qb$ and $\Qs$ as the betatron and the synchrotron tunes.   

It may be useful to note that by virtue of Eqs.~(\ref{ABS main}, \ref{bc}) the space derivatives of the two offset amplitudes, $x^+$ and $x^-$, are opposite at the bunch edges, $\partial x^{+}/\partial s=-\partial x^{-}/\partial s$ at $s=0,\,1$.

An alternative way to represent Eqs.~(\ref{ABS main}, \ref{bc}) opens if we consider the fluxes $x^+$ and $x^-$ as two parts of a single circulation $x(\psi)$ in the longitudinal phase space, with the synchrotron phase $\psi$ running from $-\pi$ to $0$ for $x^+$ part, and continuing to run from $0$ to $\pi$ for $x^-$. In other words, 
\begin{equation}
x(\psi)= \left \{
\begin{split}
x^+(s)\,, &  \;\; \mathrm{with}\; \psi=-\pi s\,,\;& -\pi \leq \psi \leq 0\;; \\
x^-(s)\,, &  \;\; \mathrm{with}\; \psi=\pi s\,, & 0 \leq \psi \leq \pi\;.
\end{split}
\right.
\label{OneXdef}
\end{equation}
This representation automatically takes into account the boundary condition~(\ref{bc}) and turns two dynamic equations~(\ref{ABS main}) into one on the circulating flux $x(\theta,\psi)$:
\begin{equation}
\frac{\pd\,x}{\pd \theta} + \frac{\pd\,x}{\pd \psi} = \frac{i q}{2}\left[ x(\psi) - x(-\psi) \right] + i\,F ,	
\label{OneXEq}
\end{equation}
with $\bar{x}=[x(\psi) + x(-\psi)]/2$. By virtue of the periodicity on the synchrotron phase $\psi$, the circulation $x$ can be expanded into a Fourier series
\begin{equation}
x(\theta,\psi)=\sum_{n=-\infty}^{\infty} A_n(\theta) \exp(i\,n\psi) \,.
\label{XFourier}
\end{equation}
After that, the problem is reduced to a set of ordinary differential equations on the time-dependent Fourier coefficients $A_n(\theta)$:
\begin{equation}
\begin{split}
& i\dot{A}_n = n A_n - \frac{q}{2}(A_n - A_{-n}) - w \sum_{m=-\infty}^{\infty} U_{nm} A_m \,, \\
& U_{nm} \equiv  {\int_0^1 ds \int_0^s ds' W(s-s') \cos(\pi ns) \cos(\pi m s')}
 \end{split}
\label{Aeq}
\end{equation}
In this form, the equations of motion can be easily solved with a proper truncation of the Fourier sums; number of calculated matrix elements $U_{nm}$ is reduced $\sim 8$~times if one notes that $U_{nm}=U_{|n||m|}=(-1)^{n-m}U_{mn}$. For those wakes when the integral $U_{nm}$ can be taken analytically, the problem can be solved at contemporary laptop with any reasonable accuracy for a negligible time.

\subsection{Eigensystem}

Without accounting for wakes, the ABS eigensystem, $x_k \propto \exp(-i \nu_k \theta)$, has been described in the original Ref.~\cite{blaskiewicz1998fast}: 
\begin{equation}
\label{EigensZeroWake}
\begin{split}
	& \nu_k = -q/2 \pm \sqrt{q^2/4 + k^2}\,; \\
	&  x^{\pm}_k(s) = C_k(\cos(k\pi s) \mp i\, \sin(k \pi s)\, \nu_k/k ) \,; 
\end{split}	
\end{equation}
for $k=0$, $\nu_0=0$, $\; x^\pm_0=C_0$, and  $\nu_k >0$ at $k>0$. Hereafter, the normalization constants $C_k$ are chosen so that the ABS eigenfunctions are of the unit norm: 
\begin{equation}
\int \limits_{-\pi}^\pi \frac{d\psi}{2\pi} |x_k|^2  =\int_0^1 ds \frac{|x^+_k|^2+|x^-_k|^2}{2}=\sum_{n=-\infty}^\infty |A_{nk}|^2=1.
\label{EigenNorm}
\end{equation}
%
It may be worth noting that the ABS no-wake eigenvalues $\nu_k$ (\ref{EigensZeroWake}) are similar to the coherent tune shifts $\omega^\mathrm{coast}_n$ of the equivalent two-stream coasting beam, $\omega^\mathrm{coast}_n=-q/2 \pm \sqrt{q^2/4+n^2 \delta \omega^2}$, with $\delta \omega$ as the revolution frequency offsets and $n$ as the conventional longitudinal harmonic number.

At zero SC, $\nu_k=k$; thus, in this simplest case the eigenfunctions $x_k(\theta,\psi)$ are just plain traveling waves, $x_k(\theta,\psi)=\exp(-ik(\theta-\psi))$, yielding standing waves for the centroid oscillations, $\bar{x}_k=\cos(\pi k s) \exp(-ik\theta)$. 

\begin{figure}[h!]
\includegraphics[width=\linewidth]{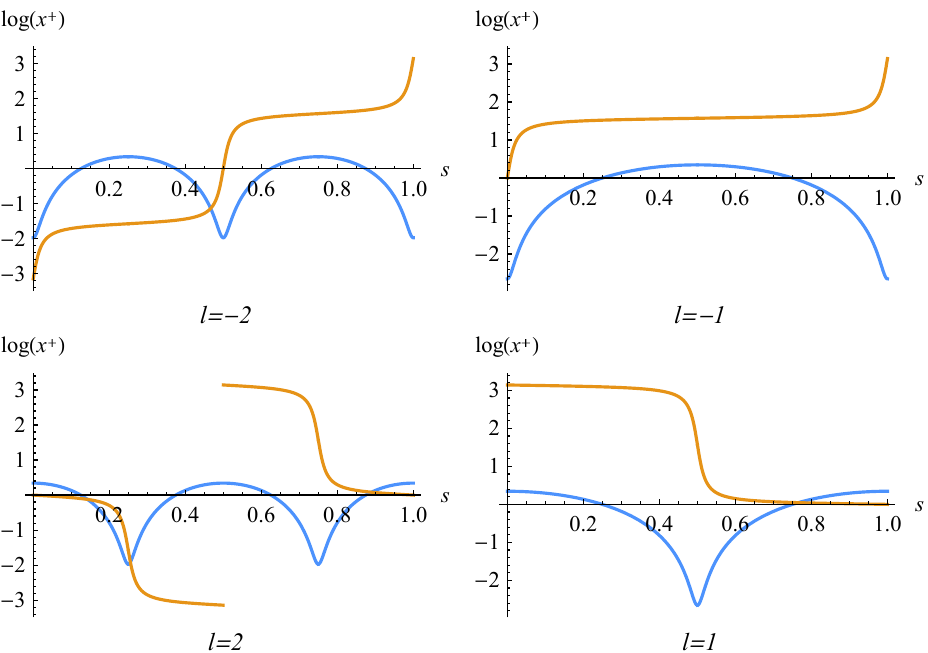}
\caption{\label{GridEigenLog_q20_w0}
	No-wake ABS eigenfunctions for strong SC, $q=20$, with their order $l$ written below each plot. The blue line shows natural logarithms of the absolute values, $\log|x^+_l(s)|$; the orange line represents the complex arguments $\arg x^+_l(s)$.   
	}
\end{figure}

\begin{figure}[h!]
\includegraphics[width=\linewidth]{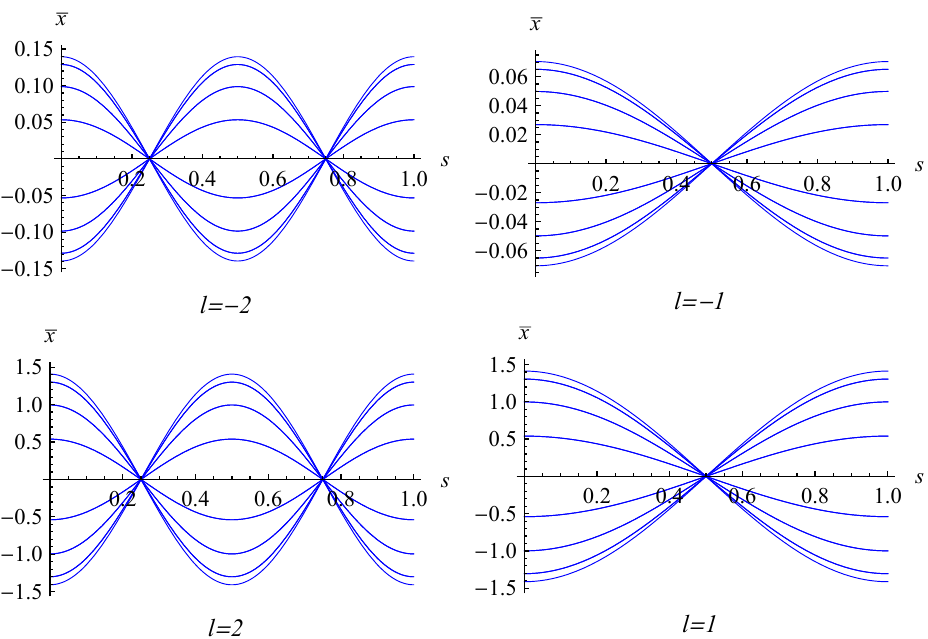}
\caption{\label{GridEigenXc_q20_w0}
	Stroboscopic snapshots of the centroid oscillations for the same case, $w=0,\, q=20$ and modes, $l=\pm1, \pm 2$, as Fig.~\ref{GridEigenLog_q20_w0} above. The opposite modes, $l$ and $-l$, show the same pattern, $\bar{x}_l(s) \propto \cos (\pi l s)$, but the amplitudes differ by a large factor $|l|/q$, reflecting almost in phase oscillations of $x^+$ and $x^-$ for the positive modes and almost out of phase ones for the negative modes.   
	}
\end{figure}

At high SC, i.e. at $q \gg 2|k|$, the low-order modes almost degenerate: $\nu_{+k} \approx k^2/q\;$, $\nu_{-k} \approx -q - k^2/q$. This spectrum tells that at high SC, the two fluxes oscillate almost identically, $x^+ \approx x^-$ for the positive modes, and they are in the opposite phases for the negative modes, $x^+ \approx -x^-$, which is indeed the case. The lowest no-wake and strong SC eigenfunctions are demonstrated in Fig.~\ref{GridEigenLog_q20_w0}. Contrary to no-SC case, the eigenfunctions $x_l$ are pretty much standing waves here; their phases $\arg x_l$ do not run, but stay constant, close to $\pi$ for the positive modes and $\pi/2$ for the negative ones, jumping by $\pi$ at the function nodes. These specific values of the phases show that the fluxes $x^+$ and $x^-$ are almost in phase for the positive modes and almost out of phase for the negative. This transfer from the traveling waves at zero SC to the standing waves at strong SC follows directly from the equation of motion~(\ref{OneXEq}) for the no-wake case, $F=0$. For zero SC, the eigenfunctions of Eq.~(\ref{OneXEq}) are ones of the translation generator $\partial/\partial \psi$, which are complex exponents, $\exp (il\psi)$. At strong SC, the equation's eigenfunctions have to be eigenfunctions of the dominating SC operator which core $\propto \delta (\psi-\psi') -  \delta (\psi+\psi')$; thus, they must have a certain parity with the phase $\psi$, being either even (positive modes and zero mode) or odd (negative modes). Since SC mixes every Fourier harmonic $n$ only with its opposite, $-n$, these eigenfunctions can be only even and odd combinations of $\exp(in\psi)$ and $\exp(-in\psi)$, i.e. they can be only $\cos(n\psi)$, for the positive modes, and $\sin(n\psi)$, for the negative ones. Figure~\ref{GridEigenXc_q20_w0} shows stroboscopic snapshots of the centroid oscillations for the same no-wake and strong SC case, as Fig.~\ref{GridEigenLog_q20_w0}, i.e. overlapping plots $\Re(\bar{x}(s)\exp(-i \theta_j))$, with the stroboscope time $\theta_j = 2\pi j/N_s$, $\; j=0,1,.., N_s-1$, and $N_s$ as an adjustable integer number. Note that with the same pattern of the identically normalized opposite modes, $l$ and $-l$, their centroid amplitudes differ at strong SC by a factor of $|l|/q \ll 1$.    
\begin{figure*}
\includegraphics[width=\linewidth]{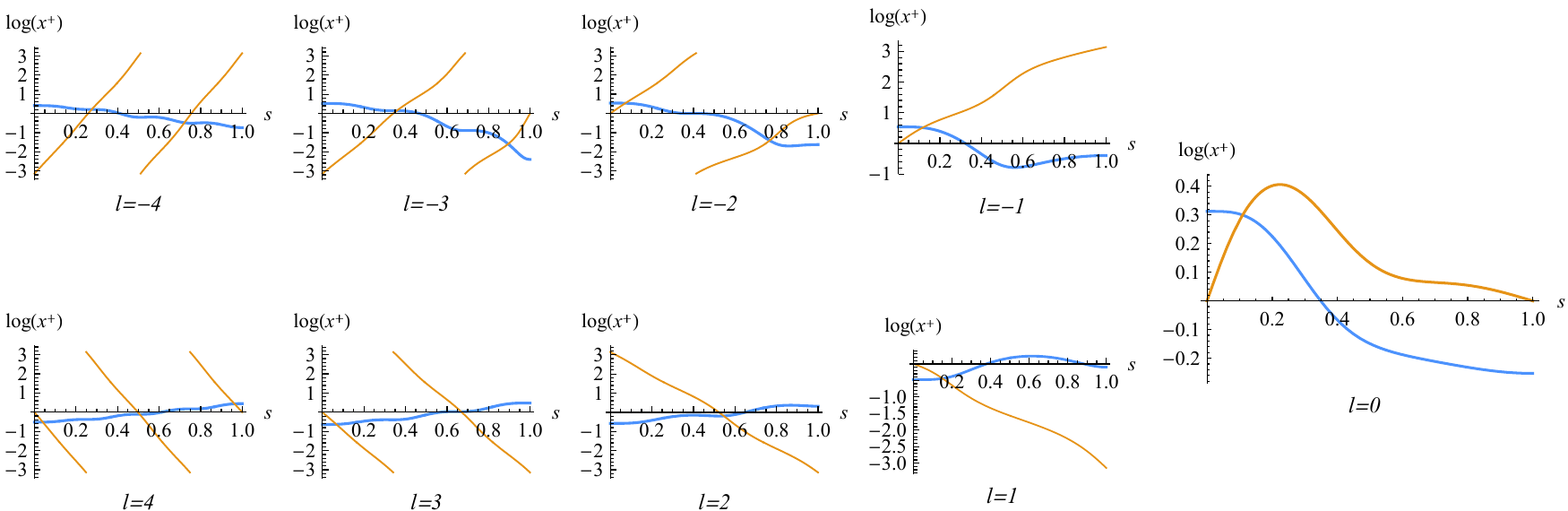}
\caption{\label{Fig:LogXQ0W13}
Same as Fig.~\ref{GridEigenLog_q20_w0}, but for no-SC and the wake parameter $w=13$, which is slightly below the TMCI threshold, $w^0_\mathrm{th}=15$, where the modes $-2$ and $-3$ couple. Note the general traveling wave pattern for all modes and that the pre-coupled negative modes are considerably head-dominated, contrary to the positive modes.
	}
\end{figure*}
\begin{figure*}
\includegraphics[width=\linewidth]{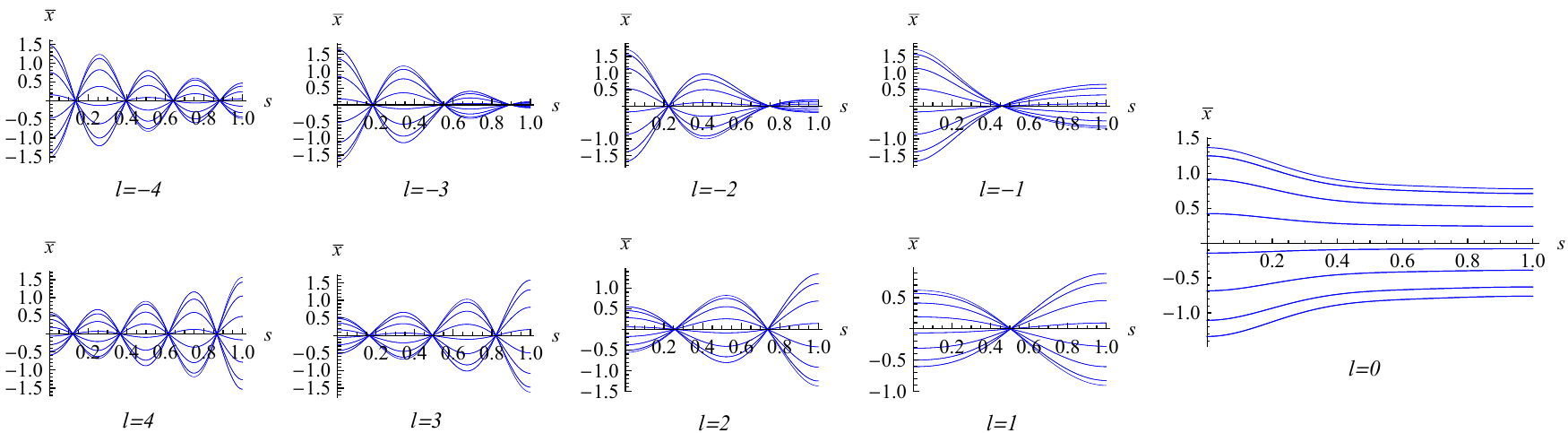}
\caption{\label{Fig:XcQ0W13}
Stroboscopic images of the centroid oscillations for the same parameters and modes as in Fig.~\ref{Fig:LogXQ0W13}. Each mode has as many nodes, as the modulus of its number.  
	}
\end{figure*}

With wake on, the eigenfrequencies $\nu_k$ shift from their no-wake values~(\ref{EigensZeroWake}), and, if the wake is large enough, they couple, giving rise to the transverse mode coupling instability, TMCI. Whatever the wake, some general features of the eigenfunctions can be stated. 
\begin{itemize}
\item{ Without any loss of generality, the eigenfunctions $x^+_k$ and $x^-_k$ at the head of the bunch can be taken as the same real number: $\Im(x^\pm_k(0))=0$.}
\item{ Due to the symmetry of Eqs.~(\ref{ABS main}) and their boundary conditions (\ref{bc}),  $x^-_k(s)=x^{+*}_k(s)$ for any eigenfunction $k$ with real eigenfrequency $\nu_k$, where $^*$ means the complex conjugate. Hence, the amplitudes are real at the tail end, $x^{\pm*}(1) = x^{\pm}(1)$.}
\item{It follows, that for the modes with real frequencies $\nu_k$ the head-to-tail phase advances $\; \mu_k$ of $x^\pm_k(s)$ are multiples of $\pi$, $\; \mu_k=\mp \pi \,k$. The centroid $\bar{x}_k(s)$ is a real function for such modes. Centroid's number of zeroes (nodes) is {\it almost always}\footnote{ The only exception known to the author to this unproven statement relates to the pre-coupled states of modes $0$ and $-1$, when the former may 'spoil' the latter, so that the mode $-1$ may be tail-dominated and be of no nodes even below the TMCI threshold}  equal to the modulus of the mode number.  
}
\end{itemize}

Following Ref.~\cite{burov2009head}, we use a term {\it strong SC}, meaning that the SC tune shift considerably exceeds all other tune shifts. In other words, it means that the SC parameter $q$ is large in comparison with the mostly involved modes' numbers $|l|$ and with the wake-driven coherent tune shifts $\simeq w U_{ll}$, i.e. $q \gg |l|$, and $q \gg w/w^0_\mathrm{th}$, where $w^0_\mathrm{th}$ is TMCI threshold value of the wake parameter at zero SC. At strong SC, the wake cannot effectively mix positive and negative modes. Thus, for strong SC, the separation between positive modes, coupled with wake, and negative modes, uncoupled with it, is effective with wake as well as without it; hence, only positive modes may play a role, i.e. the bunch longitudinal slices are rigid~\cite{burov2009head}.

\begin{figure*}
\includegraphics[width=\linewidth]{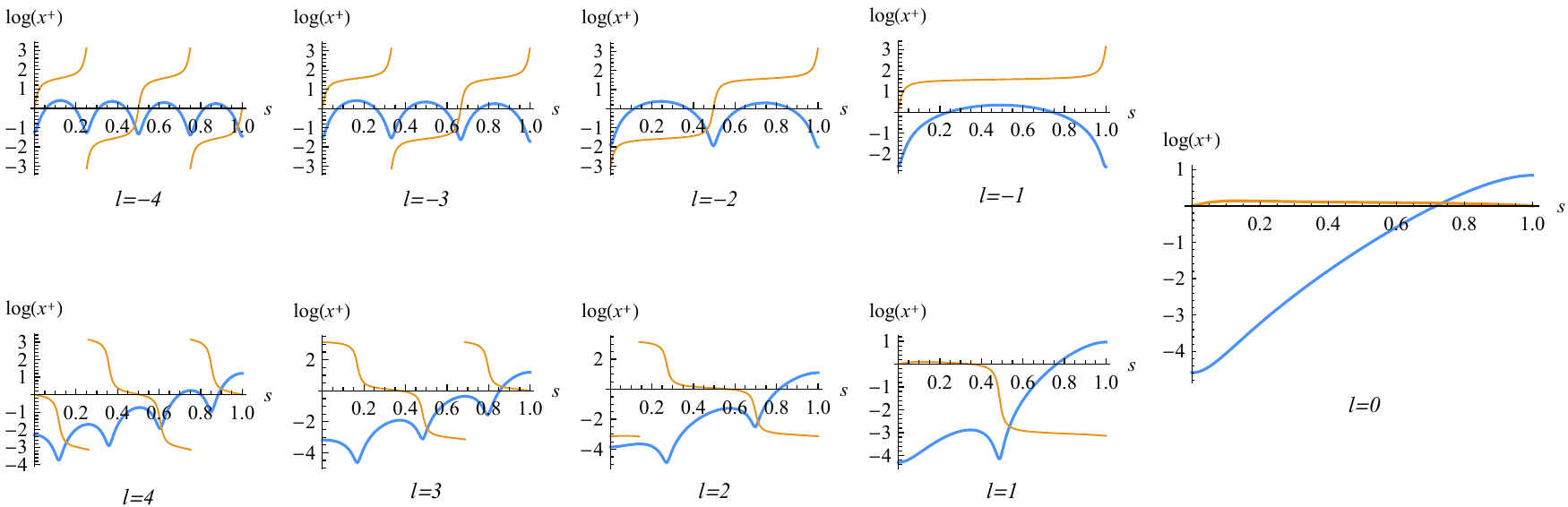}
\caption{\label{Fig:LogXQ20W13}
Eigenfunctions with the broadband resonator wake, Eq.~(\ref{BBWake}), wake and SC parameters $w=13$, $\,q=20$; compare with Figs.~\ref{GridEigenLog_q20_w0},~\ref{Fig:LogXQ0W13}. At that strong SC, the wake parameter $w$ is $\simeq 9$~times below the TMCI threshold. Blue lines show natural logarithms of the amplitudes $\log|x^+_l|$; the orange ones are reserved for the phases $\arg(x^+_l)$. All the modes are absolutely stable, $\Im \nu_l=0$, while head-to-tail amplification for the non-negative modes may exceed $100$ for these parameters; note the {\it cobra shapes}, typical for these convective instabilities. Contrary to that, the negative modes look identical to their no-wake shapes of Fig.~\ref{GridEigenLog_q20_w0}: with the out of phase motion of the $+$ and $-$ fluxes, the wake fields of the fluxes almost cancel each other.} 
\end{figure*}

\begin{figure*}
\includegraphics[width=\linewidth]{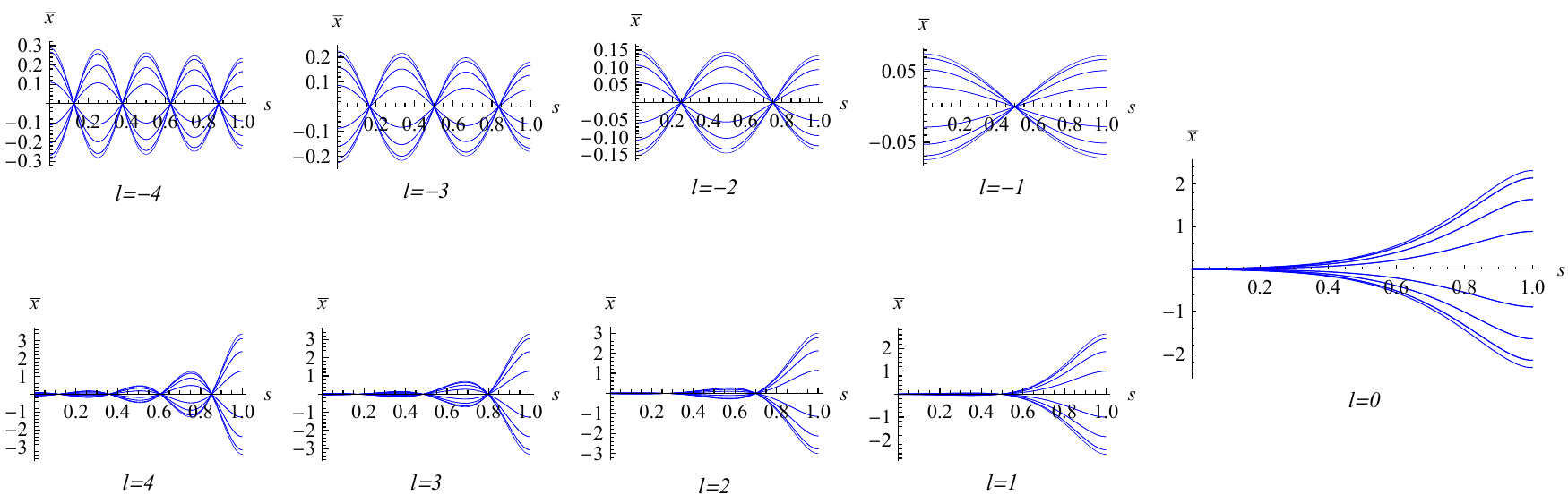}
\caption{\label{Fig:XcQ20W13}
Stroboscopic images of the centroid oscillations for the same parameters and modes as in Fig.~\ref{Fig:LogXQ20W13}. Number of nodes for each mode is identical to the modulus of its number.   
	}
\end{figure*}

Examples of the eigenfunctions, without and with SC, are presented in Figs.~\ref{Fig:LogXQ0W13}~-~\ref{Fig:XcQ20W13}  for the broadband resonator wake 
\begin{equation}
W(s)=\exp(-\alpha_r s) \sin(\bar{k}\,s)\,, 
\label{BBWake}
\end{equation}
with $\alpha_r = k_r/(2Q_r)$, $\bar{k}=\sqrt{k_r^2-\alpha_r^2}$, $Q_r=1$.  To make an accent on long bunches, especially interesting for many proton machines, a rather short resonator wake was taken, which phase advance over the bunch length $k_r=10$. The wake parameter $w=13$; it is chosen to be rather close to the no-SC TMCI threshold, $w_\mathrm{th}=15$. 

Several features of these eigenfunctions deserve to be noted: 
\begin{itemize}
\item{For the no-SC case, the phases run pretty much linearly, similarly to the no-wake no-SC case. }
\item{For the strong SC, the phases are mostly constant, quickly changing by $\pm \pi$. For the negative modes, the phases are mostly close to $\pm \pi/2$, which means that the two fluxes, $x^+$ and $x^-$, oscillate in counter-phase. For the positive modes, the phases are mostly close to $0$ or $\pm \pi$, showing that the two fluxes move together. These features are again similar to the no-wake case of Fig.~\ref{GridEigenLog_q20_w0}. }

\item{Without SC, the modes $-2$ and $-3$ couple at the wake parameter $w_\mathrm{th}=15$, just slightly above $w=13$ of Figs.~\ref{Fig:LogXQ0W13}~-~\ref{Fig:XcQ20W13}. According to the author''s observations, the pre-coupled modes are typically dominated by the head~\cite{Note1}.}
\item{At the strong SC, negative modes are not sensitive to wake, while the positive modes steeply rise to the tail, showing something like cobra shapes, with the bunch tail as the cobra head, though; see Fig.~\ref{Fig:LogXQ20W13}. The reason is that at strong SC, the two bunch fluxes oscillate almost in opposite phases for negative modes, almost cancelling their wake forces. Contrary to that, for the positive modes, the fluxes oscillate together, their wake fields add, which results in the {\it convective instability}, well seen in Figs.~\ref{Fig:LogXQ20W13},~\ref{Fig:XcQ20W13}.}
\end{itemize}
Having discussed the way the convective instabilities show themselves through the eigensystems, we may exercise another, complementary, way to look at them: the initial conditions, or Cauchy, problem.
\begin{figure*}
\includegraphics[width=\linewidth]{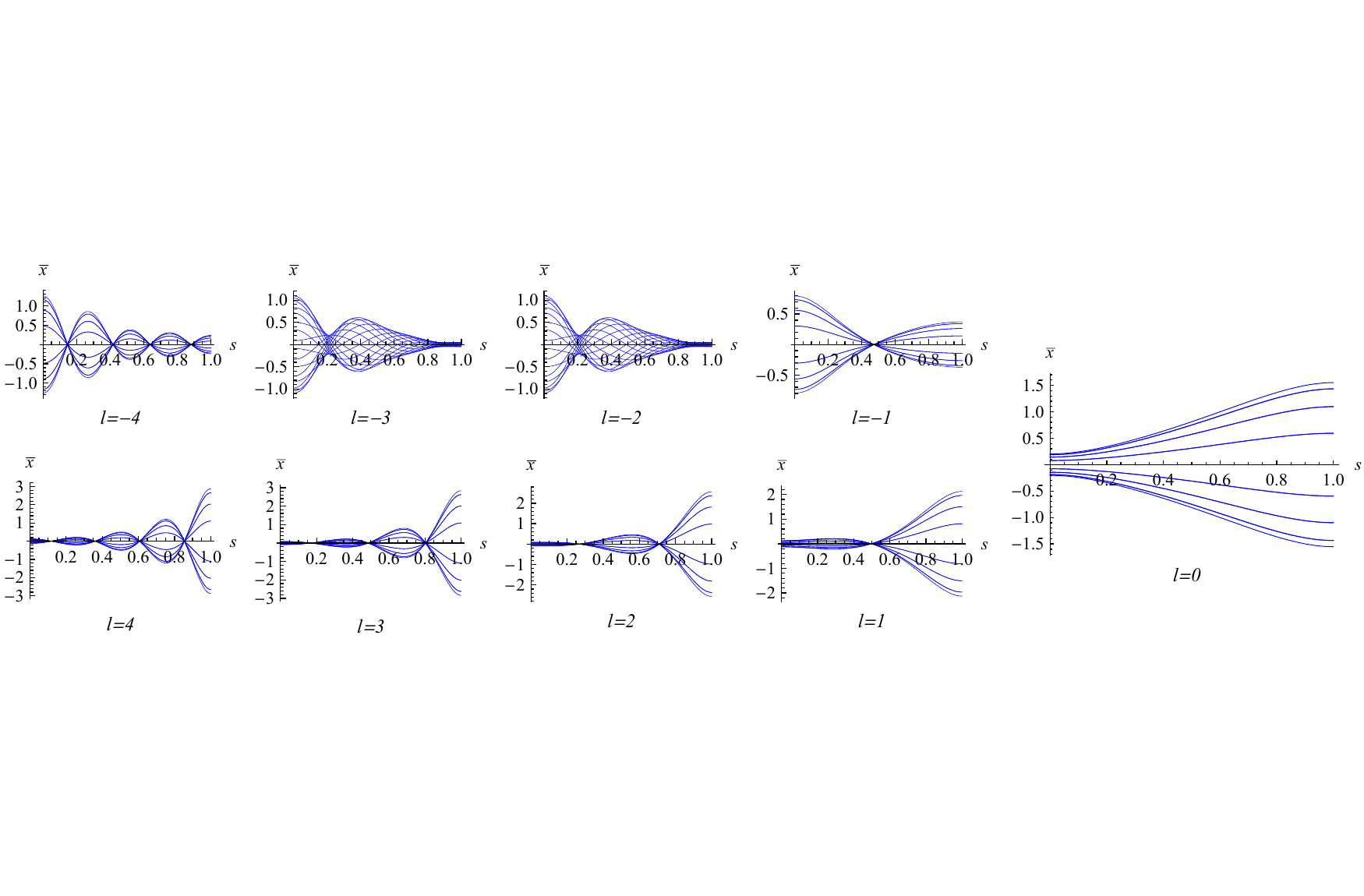}
\caption{\label{Fig:XcQ4W35}
Centroid oscillations for a moderate SC, $q=4.1$ and $w=35$, which is a bit above the TMCI threshold $w_\mathrm{th}=30$ at this SC parameter, twice as high as at zero SC. Nodes of the coupled modes $l=-2$ and $l=-3$ become waists. Note that head-dominated TMCI of the negative modes is complemented by tail-dominated SCI of the positive ones.   
	}
\end{figure*}

\section{\label{sec:ABS_Results}Cauchy problem for ABS}

\subsection{Difference Scheme}

Perhaps, the most straightforward method to solve at this point the Cauchy problem is one suggested by the Fourier form of  Eq.~(\ref{Aeq}). For the sake of diversity, though, as well as for a possibility of cross-checking, the author of this paper prefers to use another, not less effective, method of solution: a simple first-order numerical difference scheme, applied to the original equation of motion~(\ref{ABS main}) with the boundary condition~(\ref{bc}). The time derivative there can be taken as 
$$
\frac{\pd\,x}{\pd \theta} \approx (x_{\mu+1,p} - x_{\mu,p})/\Delta \theta .
$$
for both $+$ and $-$ fluxes, where the Greek characters enumerate the time steps, and the Latin ones, $p=1,2,...,P$, are used for the space; $\Delta \theta \ll 1$ and $\Delta s =1/(P-1) \ll 1$ are the time and space steps respectively. To make the algorithm numerically stable, the space derivatives have to be taken in accordance with the flux direction: 
\begin{eqnarray}
\frac{\pd\,x^+}{\pd s} & \approx & (x^{+}_{\mu ,p+1} - x^{+}_{\mu ,p})/\Delta s \,, \qquad \\
\frac{\pd\,x^-}{\pd s} & \approx & (x^{-}_{\mu,p} - x^{-}_{\mu,p-1})/\Delta s \,,
\end{eqnarray}
and the Courant condition $\Delta \theta < \pi \Delta s$ has to be satisfied. 
The boundary conditions, Eq.~(\ref{bc}), allow to express $x^{\pm}$ when the spacial indices step outside the bunch length:
\begin{equation}
\label{out}
x^{+}_{\mu ,P+1}=x^{-}_{\mu ,P-1}\,, \;\; x^{-}_{\mu ,0}=x^{+}_{\mu ,2}	 \,.
\end{equation}
With these substitutions, as well as a first-order transformation of the wake integral into a sum, the resulting equations for a $2P$-component vector $\mathbf{X}_\mu=(\mathbf{x}^+_{\mu}, \mathbf{x}^-_{\mu})$, composed of two $P$-component vectors $\mathbf{x}^{\pm}_\mu$, can be written in a matrix form 
\begin{equation}
\label{OneStepMatrix}
\mathbf{X}_{\mu+1} = \mathbf{X}_{\mu} + \Delta \theta \, \mathcal{L} \cdot \mathbf{X}_\mu , 
\end{equation}
where $\mathcal{L}$ is the time-independent infinitesimal $2P \times 2P$ generator matrix. From here, the sought-for vector at given time $\theta$ can be represented as
\begin{equation}
\label{CouchyRes}
\mathbf{X}(\theta) = \exp(\theta\,{\mathcal{L}})\mathbf{X}(0) = (\mathcal{I} +\Delta \theta \mathcal{L})^{N_\theta} \mathbf{X}(0)\,, 
\end{equation}
where $N_\theta = \theta/\Delta \theta$ is a total number of the time steps, $\mathcal{I}$ is $2P \times 2P$ identity matrix, and $\mathbf{X}(0)$ is a $2P$-vector of the initial conditions. 
The described numerical method reduces the problem to raising a $2P \times 2P$ matrix to a power $N_\theta \gg 1$. Note that this computation can be performed with only $\log_2(N_\theta)$ matrix multiplications, if the number of time steps $N_\theta$ is made an integer power of 2, making the numerical scheme extremely efficient. 

\subsection{TMCI and Convective Instability}

Let us start from the simplest example of the Heaviside step wake $W(s)=\Theta (s)$, where some analytical estimations are possible and not cumbersome, and which presents an alternative to the short broadband wake considered in the previous section. For short time intervals $\theta \ll 1$, the synchrotron motion can be neglected. Assuming the head-tail amplification coefficient $K$ to be large enough, its natural logarithm can be estimated:    
\begin{equation}
\log K \simeq 2(i w \theta)^{1/2} \,. 
\label{Kamp}
\end{equation}
Hence, the maximally achievable amplification scales as 
\begin{equation}
\log K \propto  \sqrt{w} \,.
\label{KampMax}
\end{equation}
This statement can be checked by means of the described solution of the Cauchy problem for the ABS model. Without SC, its TMCI threshold is $w_\mathrm{th}=1$, see Refs.~\cite{blaskiewicz1998fast, PhysRevAccelBeams.21.104201}. 

\begin{figure}[h!]
\includegraphics[width=\linewidth]{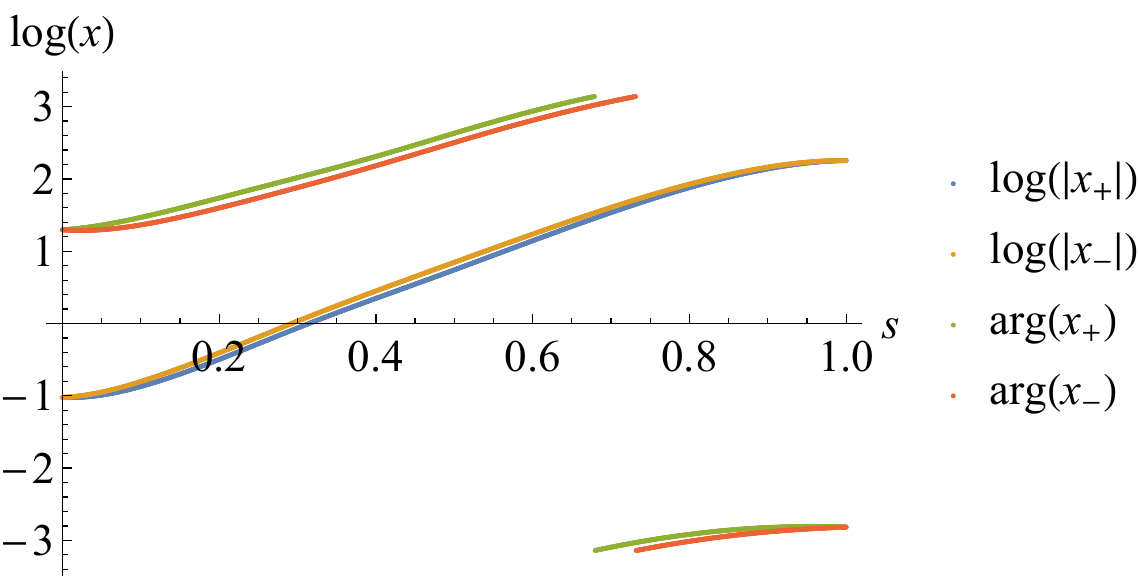}
\caption{\label{Fig:CauchyThetaChi1}
	Evolution of constant initial offset $x^{\pm}(s)=1$ after 1.5 synchrotron periods, with the SC parameter $q=20$, step wake at the no-SC threshold, $w=1$.  Natural logarithms of the absolute values and complex arguments of the amplitudes $x^{\pm}$ are shown. Note that the complex amplitudes of the fluxes are almost identical, $x^+ \approx x^-$.
	}
\end{figure}

\begin{figure}[h!]
\includegraphics[width=\linewidth]{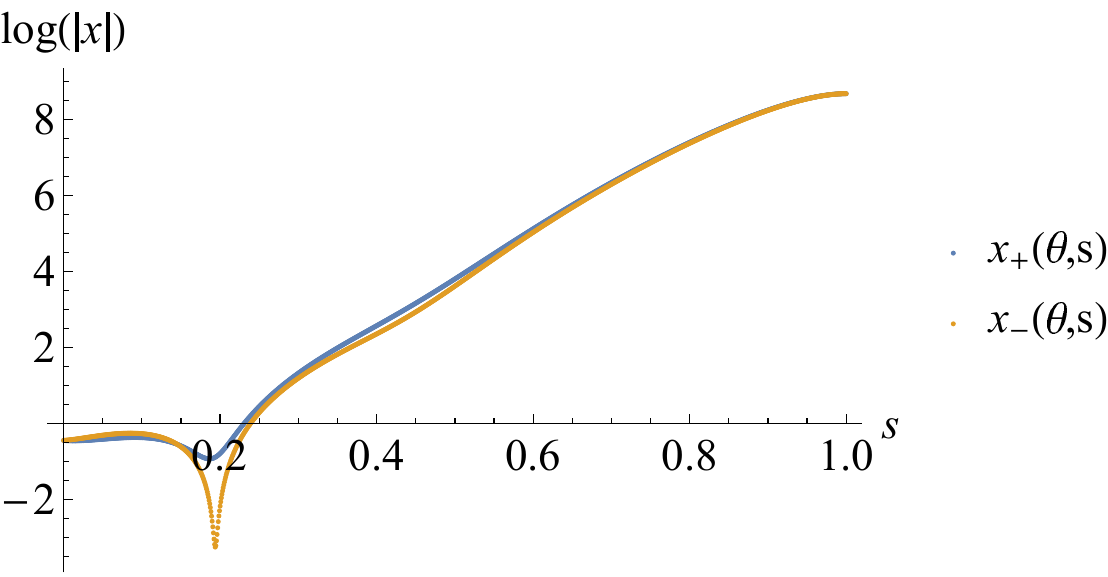}
\caption{\label{Fig:CauchyThetaChi20}
Same as the previous figure, but with 20 times larger wake, $w=20$. With that high amplification, the oscillations are still absolutely stable, as they must be, since the wake is below its threshold value. Note the cobra shape, typical of the SCI.  
	}
\end{figure}

Figures~\ref{Fig:CauchyThetaChi1},~\ref{Fig:CauchyThetaChi20} show the results of evolution after 1.5 synchrotron periods of the constant initial offset $x_{\pm}(s)=1$; the wake parameters are $w=1$ and $w=20$ correspondingly, and the SC parameter $q=20$. It has been separately checked that the amplification reaches its limit after $\sim$1 synchrotron period, so for both cases the bunch is stable in the absolute sense, notwithstanding its wake being 20 times above the no-SC threshold for the latter case. Several things are worth mentioning for these plots:
\begin{itemize}
\item{At the no-SC threshold, $w=1$, the convective amplification is already significant, $K \simeq 10$. }
\item{The two fluxes expectably oscillate in phase: out of phase oscillations are detuned by the SC tune shift from the wake-coupled motion of the centroid $(x^+ +x^-)/2$. That is why SC boosts head-to-tail signal amplification, making the bunch longitudinal slices rigid.}
\item{The convective instability leads to the acclivitous cobra shape of the amplitudes $|x^\pm|$, with zero derivative at the bunch tail.}
\item {While the wakes differ by a factor of 20, the logarithms of the amplification confirm the scaling~(\ref{KampMax}), showing that they differ by a factor close to $\sqrt{20}$.}
\item {These plots make it possible to fit the numerical factor for the amplification (\ref{KampMax}) for the ABS model with the step-like wake:  $\log K \simeq 2 \sqrt{w}$.}
\end{itemize}
After this brief examination of the theta-wake, let us come back to the broadband wake, Eq.~\ref{BBWake}, with the same phase advance $k_r=10$ and the quality factor $Q_r=1$ as above, to compare the complementary results of the eigensystem problem and the Cauchy problem with constant initial condition, $x^\pm=1$, for that physically interesting case.

\begin{figure}
\includegraphics[width=\linewidth]{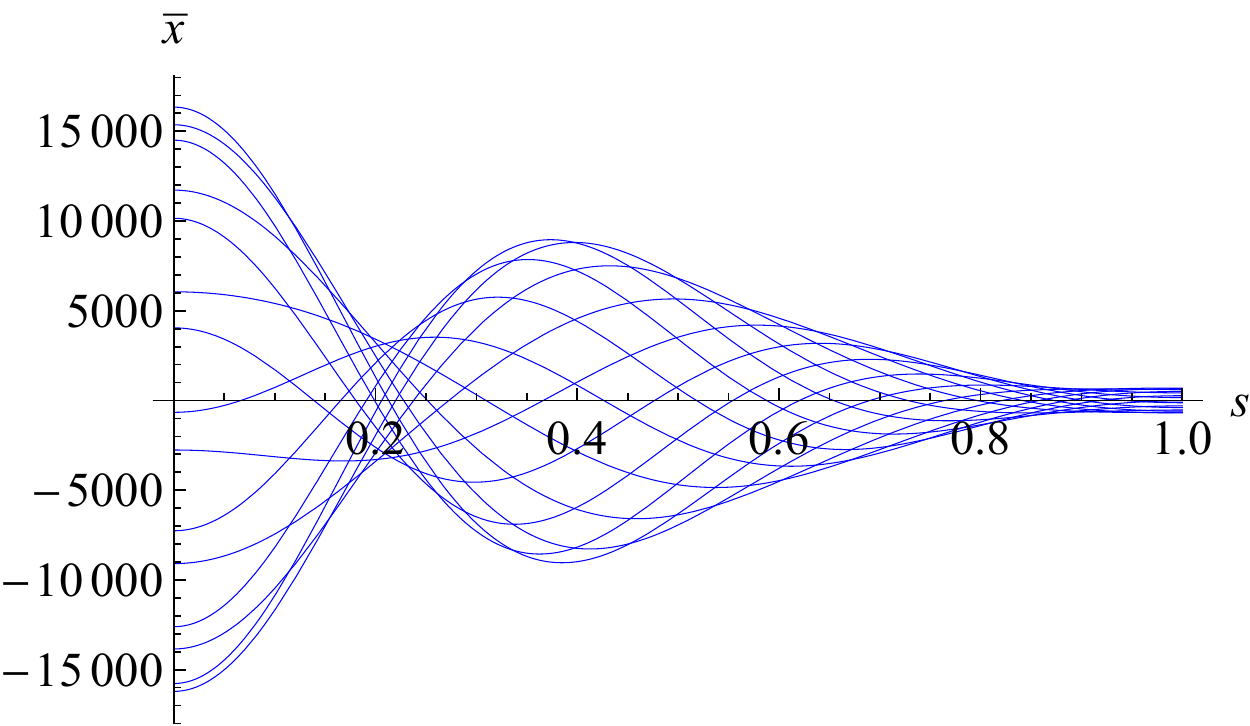}
\caption{\label{Fig:XStroboQ4W35Time8Ts}
	Evolution of the standard initial conditions $x^\pm=1$ after $8$ synchrotron periods for the same wake and SC parameters as in Fig.~\ref{Fig:XcQ4W35}, $q=4.1$ and $w=35$, slightly above the TMCI threshold wake value $w_\mathrm{th}=30$ at this SC, twice as it is at no SC case. Identity of this pattern with the coupled eigenfunctions $l=-2$ and $l=-3$ of Fig.~\ref{Fig:XcQ4W35}  serves as a good cross-check.
	}
\end{figure}

Figure~\ref{Fig:XStroboQ4W35Time8Ts} demonstrates evolution of the standard initial conditions $x^\pm=1$ after $8$ synchrotron periods for the same wake and SC parameters as in Fig.~\ref{Fig:XcQ4W35}, $q=4.1$ and $w=35$, slightly above the TMCI threshold wake value $w_\mathrm{th}=30$ at this SC, twice as it is at no SC case.  Identity of this pattern with ones of the coupled eigenfunctions $l=-2$ and $l=-3$ of Fig.~\ref{Fig:XcQ4W35}  serves as a good cross-check.

\begin{figure}[h!]
\includegraphics[width=\linewidth]{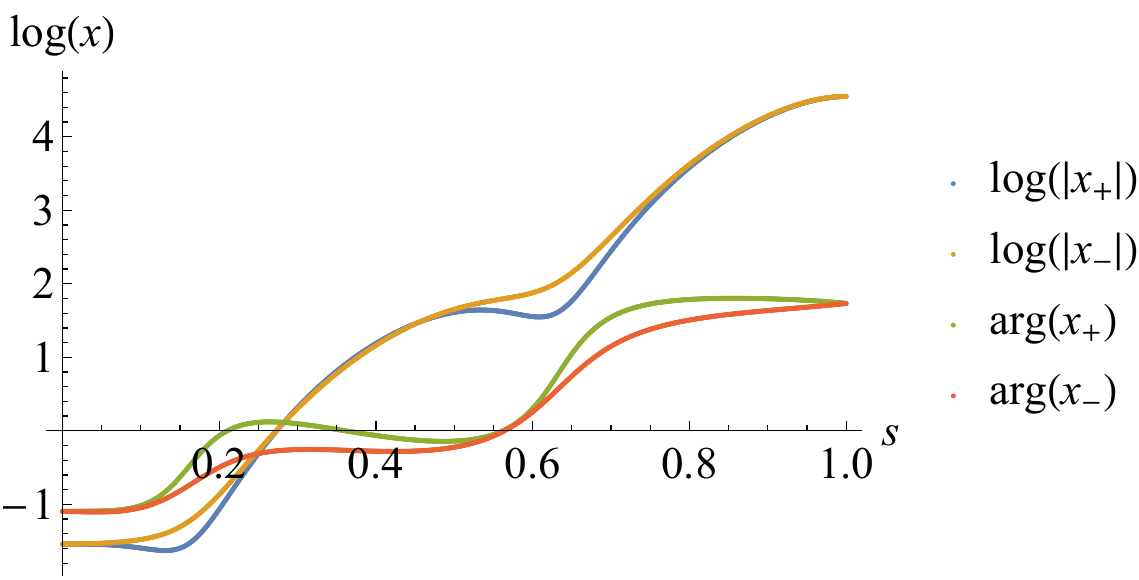}
\caption{\label{Fig:CauchyKr10Chi13Qsc20}
	Evolution of the standard initial perturbation after $1.5$ synchrotron periods for the wake parameter $w=13$ and strong SC, $q=20$. Note that the two amplitudes are close; compare with Fig.~\ref{Fig:LogXQ20W13}.
	}
\end{figure}

\begin{figure}[h!]
\includegraphics[width=\linewidth]{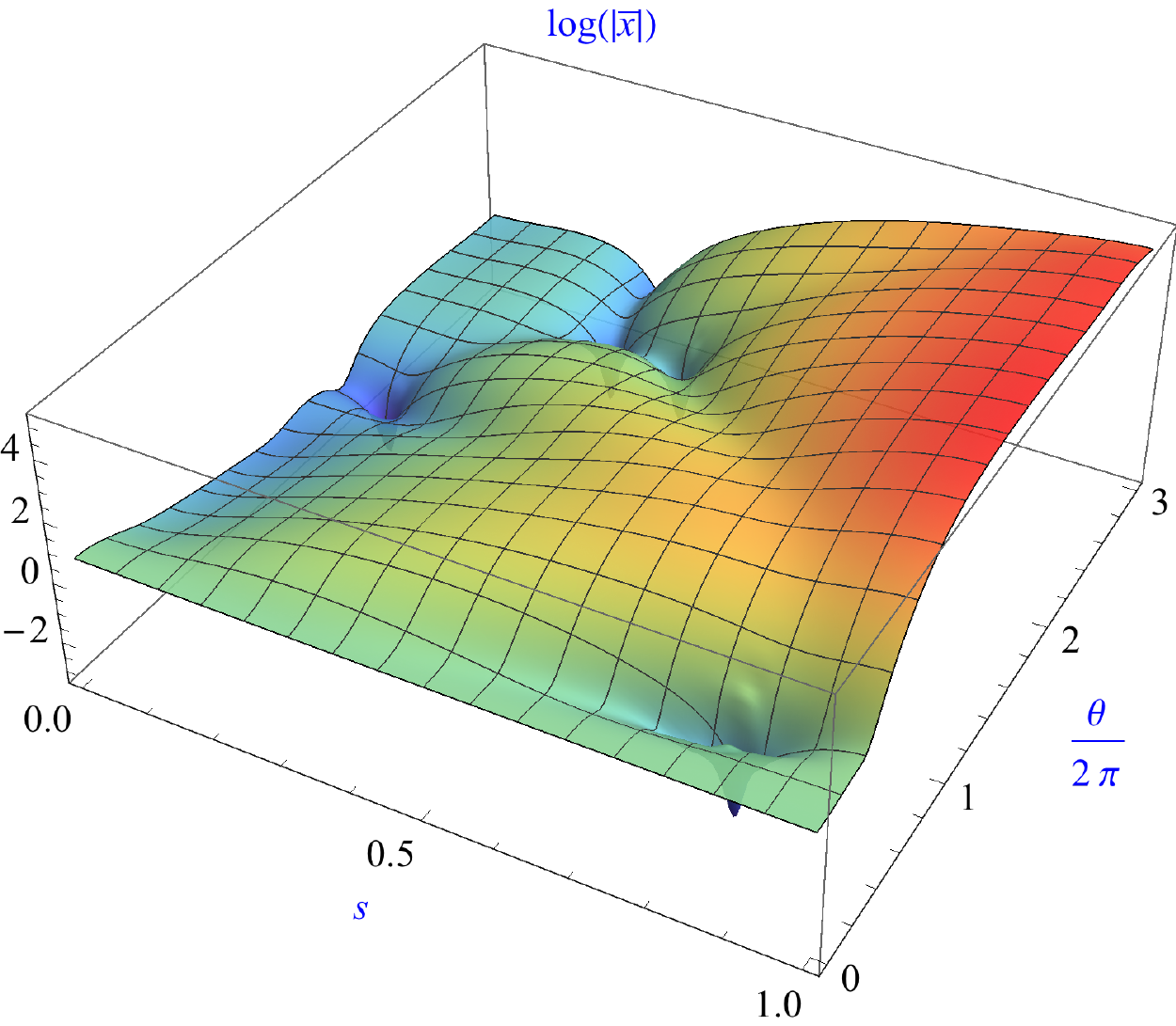}
\caption{\label{Fig:3DConv}
	Time evolution of the local centroids $\bar{x}(\theta,\,s)=[x^+(\theta,\,s) + x^-(\theta,\,s)]/2$ for the same case, i.e. for  $q=20\,,\; w=13$ and constant initial conditions, $x^\pm=1$. The amplification is saturated within $\sim$1 synchrotron period. 
	}
\end{figure}

\begin{figure}[h!]
\includegraphics[width=\linewidth]{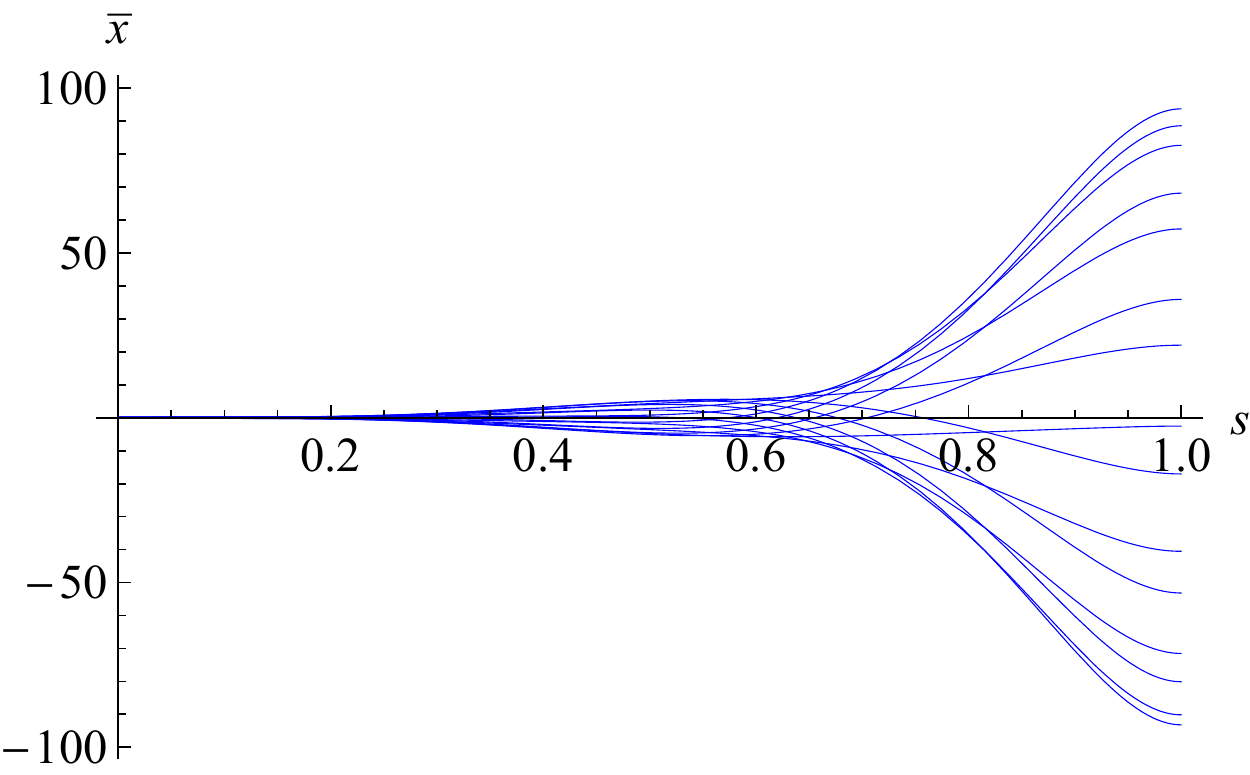}
\caption{\label{Fig:XbarStrobo}
	Stroboscopic image of the beam centroid for the same parameters as in Fig.~\ref{Fig:3DConv} after $1.5$ synchrotron periods. Note that there are no nodes.}
\end{figure}

Figure~\ref{Fig:CauchyKr10Chi13Qsc20} shows what happens at $w=13$ and $q=20$ with the initial perturbation, $x^\pm=1$, after $1.5$ synchrotron periods; the amplification coefficient can be compared with such of zero mode of the related Fig~\ref{Fig:LogXQ20W13}. The convective instability saturation is demonstrated by the 3D plot of Fig~\ref{Fig:3DConv}. Contrary to the absolute instabilities, for the convective ones there is no selection with time of the most unstable mode, since all the modes are stable in the absolute sense, all the growth rates are zeroes. That is why the practically dominating constant initial perturbation excites several modes, and none of them is going to be stressed at the following evolution. As a results, the nodes of one convectively unstable mode overlap with antinodes of the neighbor modes, excited by the same initial perturbation, which smears all the nodes. Thus, no nodes have to be observed for the convective instability, unless a special mode is carefully excited at the beginning. This statement is illustrated by Fig.~\ref{Fig:XbarStrobo}, showing the centroid stroboscopic plot after $1.5$ synchrotron periods for $q=20$ and $w=13$, i.e. for the same conditions as in Fig.~\ref{Fig:3DConv}.    

\begin{figure}[h!]
\includegraphics[width=\linewidth]{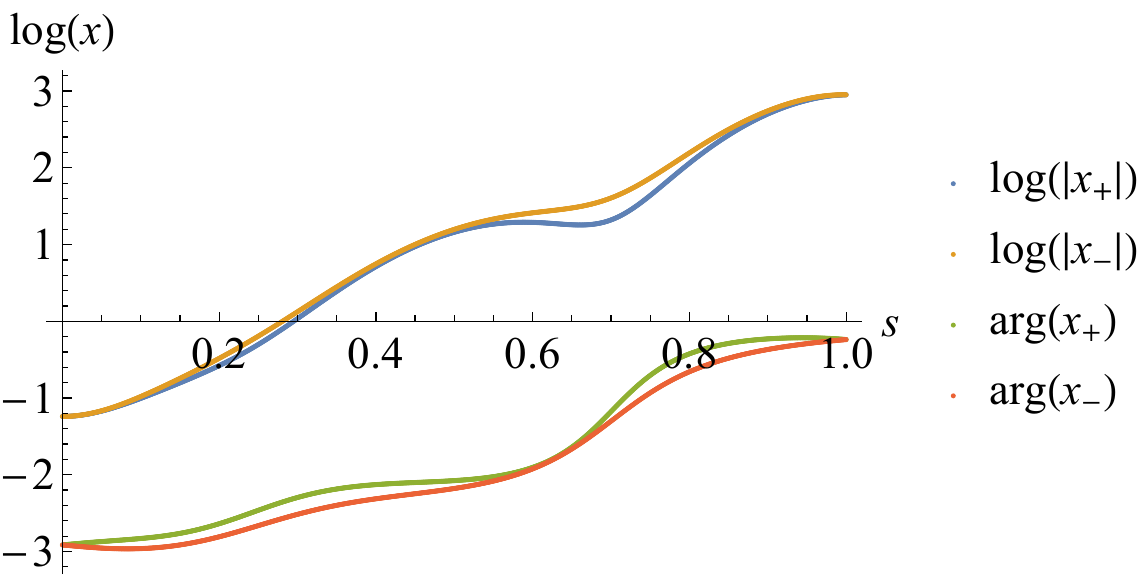}
\caption{\label{Fig:LowChiK}
	Same as Fig.~\ref{Fig:CauchyKr10Chi13Qsc20}, but with weaker wake $w=7$, i.e. 2 times lower than the no-SC TMCI threshold. Thus, with strong SC the bunch may get considerably more unstable than without it.
	}
\end{figure}

Since strong SC makes the bunch slices rigid, and thus, maximally coupled with wake, with strong SC the bunch may get considerably more unstable than without it; this is demonstrated by Fig.~\ref{Fig:LowChiK} showing significant convective amplification at $q=20$ and $w=7$, i.e. for the wake parameter of half the no-SC TMCI threshold.

%

%
%
%
%

\subsection{Absolute-Convective Instability}

The considerations and examples above demonstrate one important thing. Although at strong SC TMCI {\it vanishes}, it does not mean that in reality the beam becomes much more stable: the amplification of the saturating convective instability, SCI, can be intolerably large already at the wake parameter corresponding to the no-SC TMCI threshold, if not below that, so SCI may well be not any less dangerous than the TMCI. Moreover, the SCI, dangerous by itself, opens a door for one more type of instability. With high convective amplification, even a weak tail-to-head feedback by means of a multibunch or over-revolution wake, negligible by itself, may be sufficient to make the beam unstable in the absolute sense. The convective instability may work as a huge amplification of the otherwise insignificant mechanism of an absolute instability. In this respect, the convective instability constitutes a sort of fragile metastable state. The absolute instability generated by such amplification may be called {\it absolute-convective instability}, or ACI. 

\begin{figure}[h!]
\includegraphics[width=\linewidth]{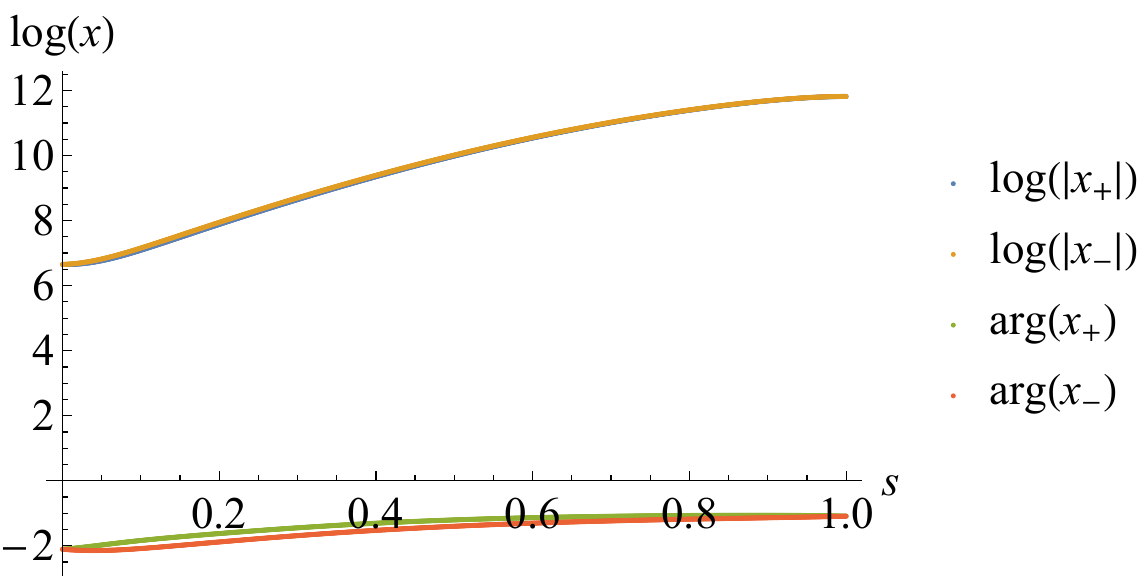}
\caption{\label{Fig:CauchyKr10Chi15Qsc20GainTm1}
	Evolution of the constant initial conditions $x^\pm(s)=1$ after time $\theta= 32 \cdot 2\pi$, or 32 synchrotron periods, with the gain so small that $g\, \theta =1$. The growth rate is $\sim$6 times higher than what the gain provides by itself. The wake phase advance $k_r=10$, the SC parameter $q=20$, the wake parameter corresponds to the no-SC TMCI threshold, $w=15$.    
	}
\end{figure}

The simplest way of modeling the tail-to-head over-revolution wake is to add to both right-hand-sides of Eqs.~(\ref{ABS main}) a center-of-mass anti-damper term $g \int_0^1 \dd s \bar{x}(s)$, where $g$ is the gain. Taken by itself, this term would drive an instability with the growth rate $g$ per synchrotron radian, or $2\pi g$ per synchrotron period. With the convective instability, the absolute growth rate, caused by the same gain, can be significantly larger. An example of such a dramatic amplification of the growth rate is shown in Fig.~\ref{Fig:CauchyKr10Chi15Qsc20GainTm1}.  For this case, the growth rate is $\sim$6 times higher than the anti-damper gain would provide alone. Note that the ACI looks similar to SCI, having alike cobra shape and rigid slices.  
\begin{figure}[h!]
\includegraphics[width=\linewidth]{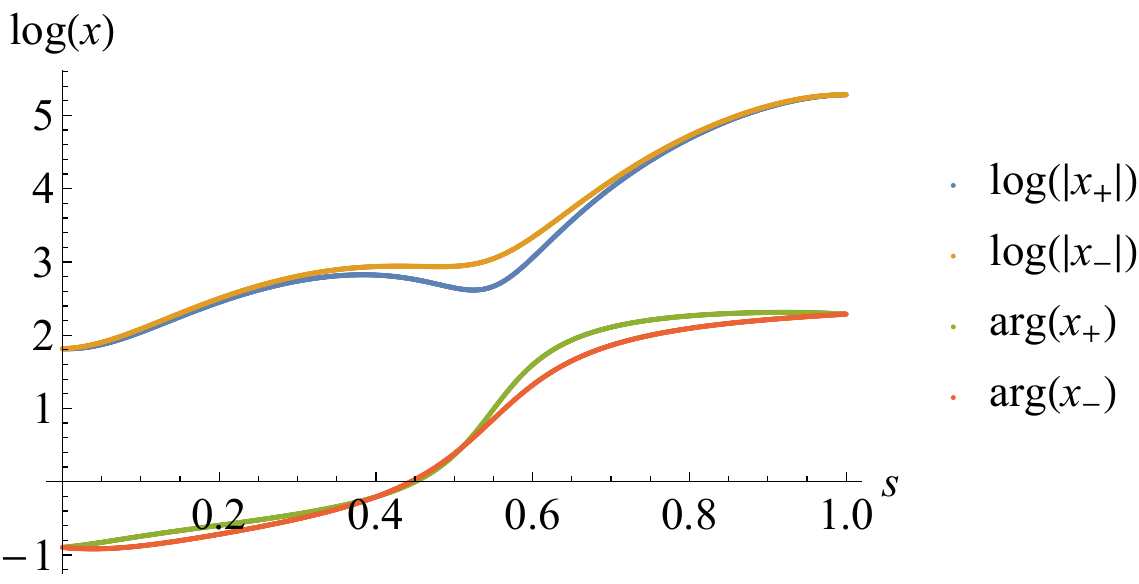}
\caption{\label{Fig:DACI}
	ACI driven by damping ({\it sic!}), with the gain $g=-0.024=-0.15/T_s$, for the wake parameter $w=7$ and SC parameter $q=20$. Evolution of the initial constant offset $x^\pm=1$ is shown after 10 synchrotron periods. Pure convective instability, SCI, for these wake and SC parameters is shown in Fig.~\ref{Fig:LowChiK}.    
	}
\end{figure}
\begin{figure}[h!]
\includegraphics[width=\linewidth]{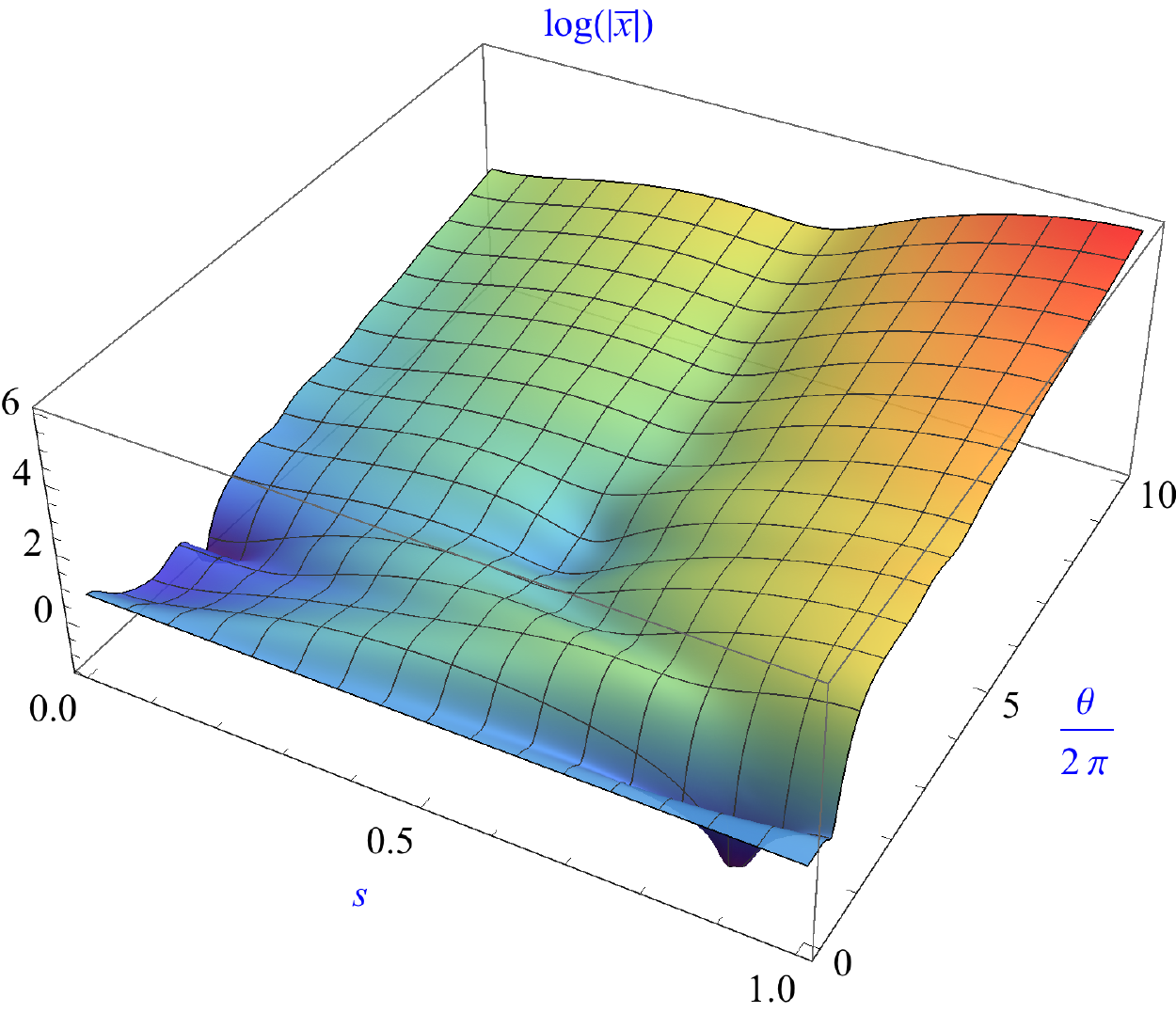}
\caption{\label{Fig:DACI3D}
	Time evolution of the ACI for the same parameters as Fig.~\ref{Fig:DACI}. An exponential growth is clearly seen.  
	}
\end{figure}

At this point, one may ask the following. If convective instabilities amplify external anti-damping, turning it into a much faster ACI, wouldn't they enhance external damping as well, making its effect even more stabilizing? Well, the answer is worse than a simple no. In fact, the convective instability turns any damper, with whatever phase, into an ACI generator. This statement deserves to be doubly stressed, since it may seem counter-intuitive: yes, even a normal bunch-by-bunch resistive damper works as an ACI generator, even for moderately amplified convective instability, considerably below the no-SC threshold; Figs.~\ref{Fig:DACI} and \ref{Fig:DACI3D} present an example. To be more precise, it may be said that the damper is just useless, if its gain is too small; then, with a higher gain, the damper shows itself as an enhancer of the SCI, and at a slightly higher gain the damper triggers the absolute-convective instability. Qualitatively this sequence of stages is the same for all gain phases, although the ACI threshold shows some quantitative dependence on this phase. The reason for this detrimental effect of any center-of-mass damper can be seen in the properties of the non-negative modes at strong SC, presented in Fig.~\ref{Fig:LogXQ20W13}. Due to the cobra shapes of the modes, the damper sees only their tails, acting back on the whole bunch proportionally to the tail offset. However, the tail motion is in fact driven by the head, which phase differs by $l\pi$ from the tail one, where $l$ is the mode number. Thus, whatever the gain phase, either even or odd modes will get a positive feedback. As a result, for conventional resistive damper, an odd positive mode with the largest coupling with the wake will be most ACI-unstable. For the broadband wake example, presented in Fig.~\ref{Fig:LogXQ20W13}, it is the mode $l=1$. Indeed, this very mode can be recognized in the ACI evolution presented in Fig.~\ref{Fig:DACI}, where the head-tail phase difference is about $\pi$.

\section{\label{sec:NT} Why Wrong Worked Right}

\begin{figure}[h]
\includegraphics[width=\linewidth]{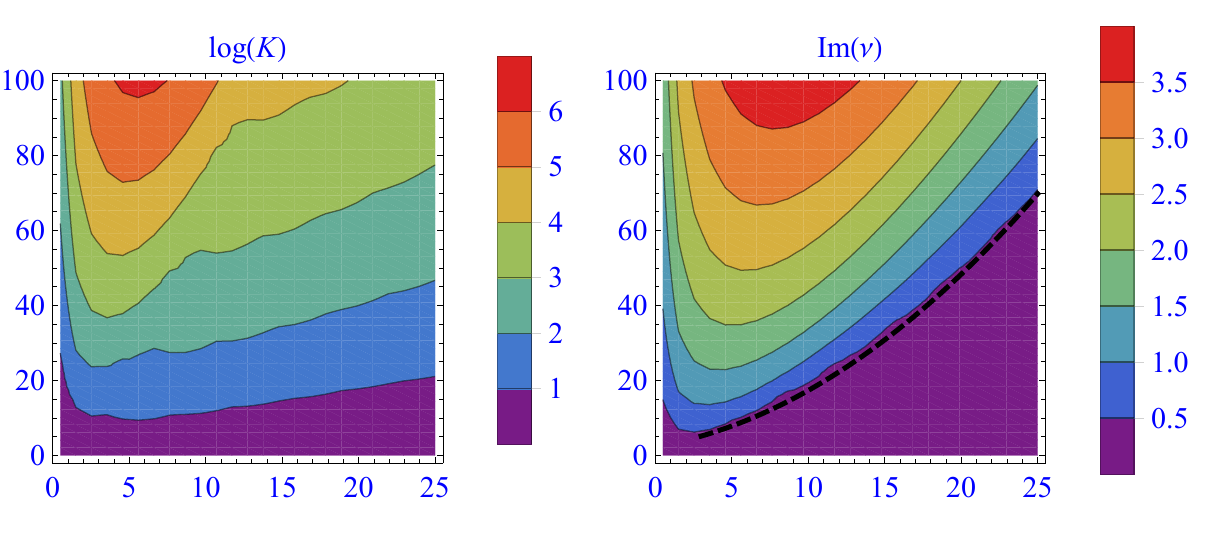}
\caption{\label{ContourKRateq0}
	Left: Contour plot for natural logarithm of the maximal head-to-tail amplification $\log K$ versus wake phase advance $k_r$, horizontally, and its amplitude parameter $w$, vertically, for the broadband case, $Q_r=1$, and no SC. Right: TMCI growth rate for the same parameters; the black dashed line is the no-SC TMCI threshold, Eq.~(\ref{NoSCTMCI}), according to Ref.~\cite{Burov:2018iid}. 
	}
\end{figure}

\begin{figure}[h!]
\includegraphics[width=\linewidth]{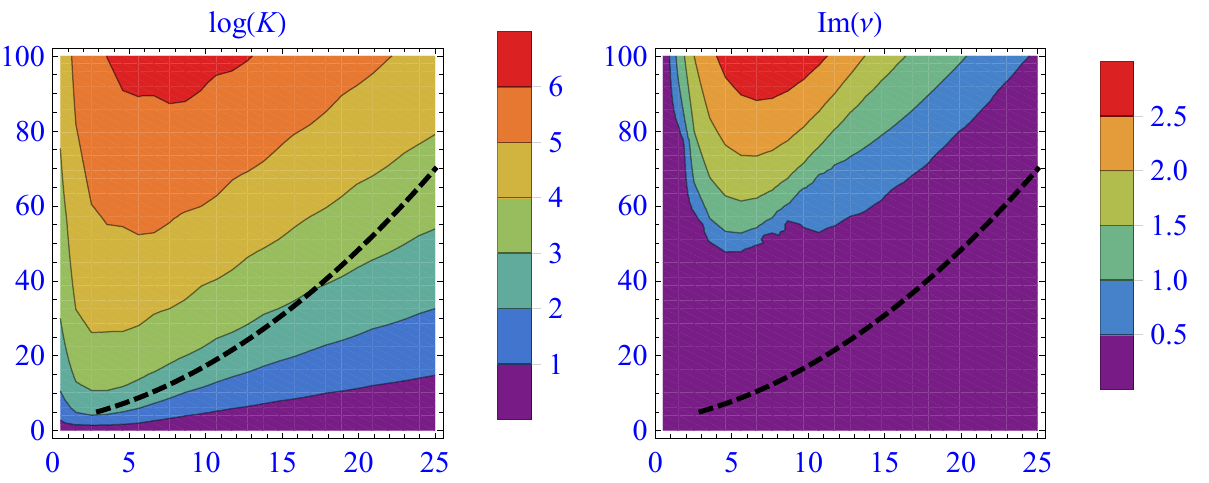}
\caption{\label{ContourKRateq5}
	Same as Fig.~\ref{ContourKRateq0}, for SC $q=5$.  The TMCI threshold moves up with SC. The black dashed line is the same no-SC TMCI threshold, Eq.~(\ref{NoSCTMCI}) . 
	}
\end{figure}

\begin{figure}[h!]
\includegraphics[width=\linewidth]{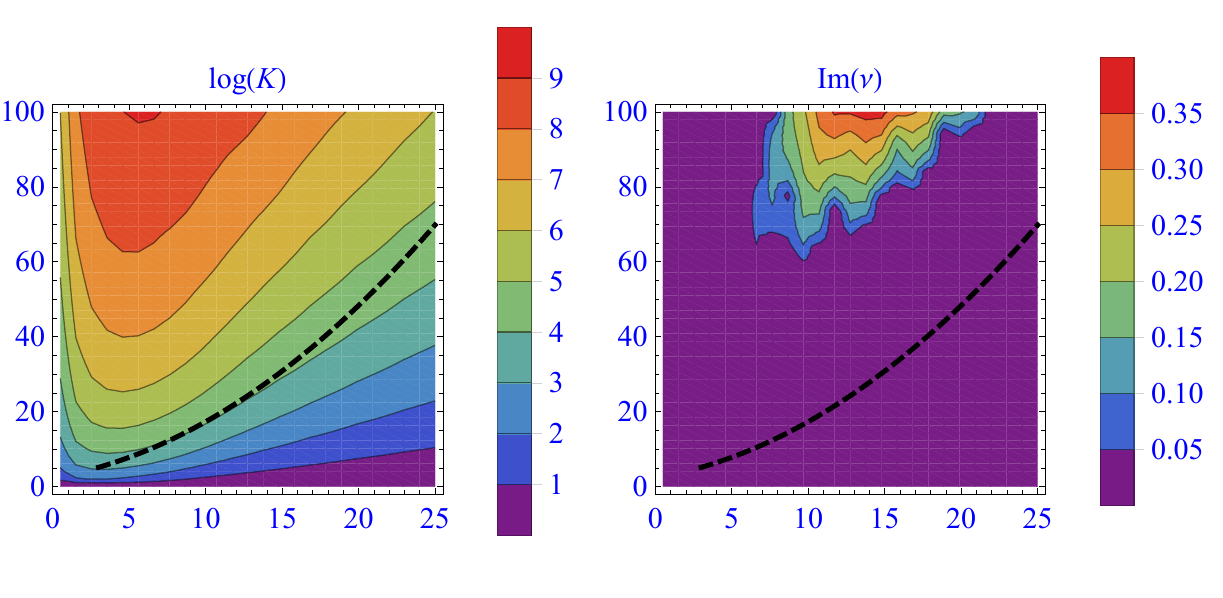}
\caption{\label{ContourKRateq10}
	The same, for larger SC, $q=10$. While TMCI threshold recedes, the amplification grows.   
	}
\end{figure}

\begin{figure}[h!]
\includegraphics[width=\linewidth]{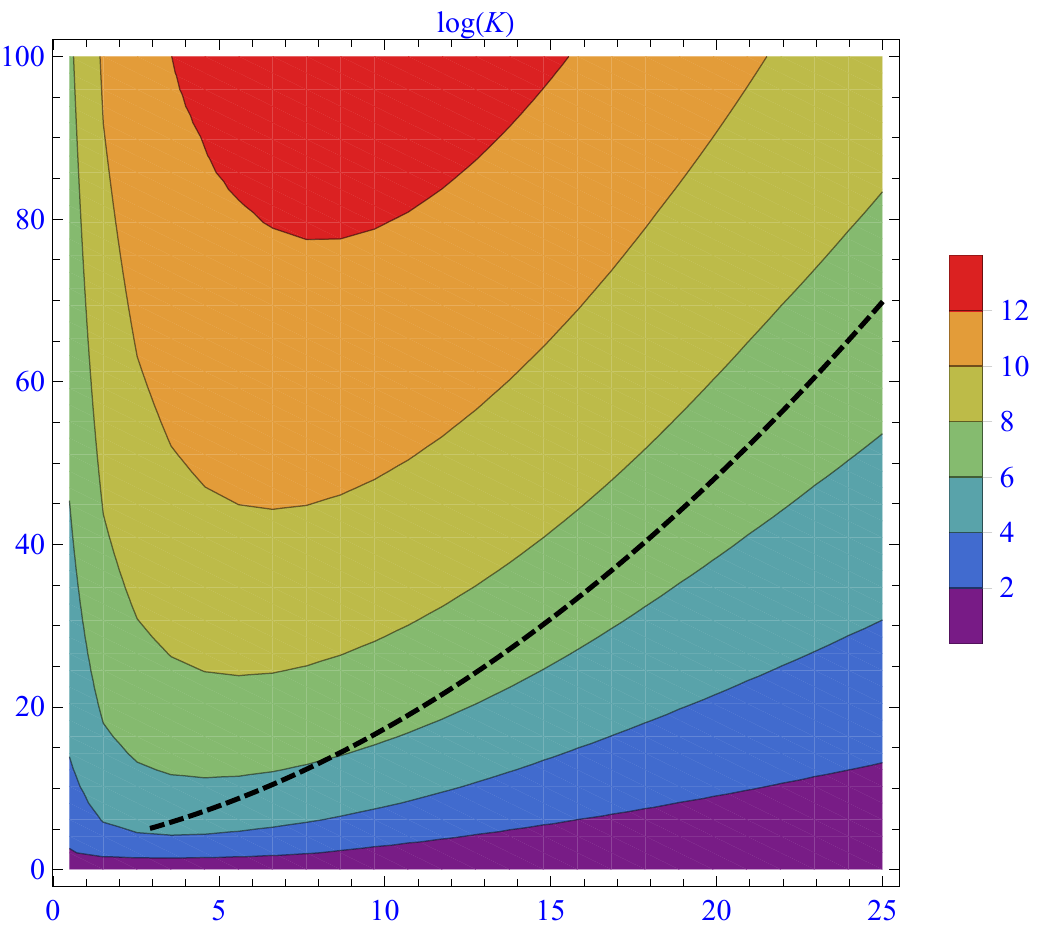}
\caption{\label{ContourKq20}
	Amplification for SC $q=20$. The black dashed line of no-SC TMCI threshold is close to the contour line $K \simeq 300 - 1000$ for large interval of the phase advances. For the entire area of the parameters, the system is absolutely stable, $\Im \nu =0$. 
	}
\end{figure}

According to Ref.~\cite{Bartosik:2013qji}, the intensity threshold at CERN SPS is fairly well described, both at the old Q26 and new Q20 optics, by no-SC TMCI threshold formula, presented therein as its last Eq.~(C.207), with a reference to~\cite{Metral:2004vi}. A recent derivation of this formula with a discussion of its numerical factor, for ABS and Gaussian bunches can be found in Ref.~\cite{Burov:2018iid} for no-SC. For the ABS with the broadband wake, $Q_r=1$, this threshold can be written as
\begin{equation}
w_\mathrm{th}^0 =2.3+0.7\,k_r + 0.08\,k_r^2\,.  
\label{NoSCTMCI} 
\end{equation} 
While at new SPS optics SC cannot be considered to be really strong, at the old Q26 case it was strong indeed; in terms of the ABS model, its SC parameter with Q26 can be estimated as $q \approx 0.5 \max \Delta Q_\mathrm{sc}/Q_s \approx 20$.
Reasonably well guidance, provided by the no-SC formula at strong SC case should not be possible according to unanimous claims of the theoretical works~\cite{blaskiewicz1998fast, burov2009head, PhysRevSTAB.14.094401}, so the question is why did this happen? A brief  answer to this question was already suggested in several parts of this paper: the same formula may suggest reasonable estimations for both no-SC TMCI threshold and the bunch intensity limit imposed by too high convective amplification at strong SC. In this section, this statement is illustrated by special computations, presented in Figs.~\ref{ContourKRateq0}--\ref{ContourKq20}. These figures show contour plots of the convective amplification and TMCI growth rates for various SC and wake parameters. For each given set of parameters, we compute two complementary maxima over the collective modes $l$, with their tunes $\nu_l$ and eigenfunctions $x_l(\psi)$: the maximal growth rate, $\Im \nu \equiv \max_l \Im(\nu_l)$, and the maximal amplification 
\begin{equation}
K \equiv \max_l \left| \frac{x_l(\pi)}{x_l(0)} \right| =\max_l \left| \frac{ \sum_{n=-\infty}^{\infty} (-1)^n A_{nl}}{ \sum_{n=-\infty}^{\infty} A_{nl}} \right|\,. 
\label{LogK}
\end{equation} 
\begin{figure}[h!]
\includegraphics[width=\linewidth]{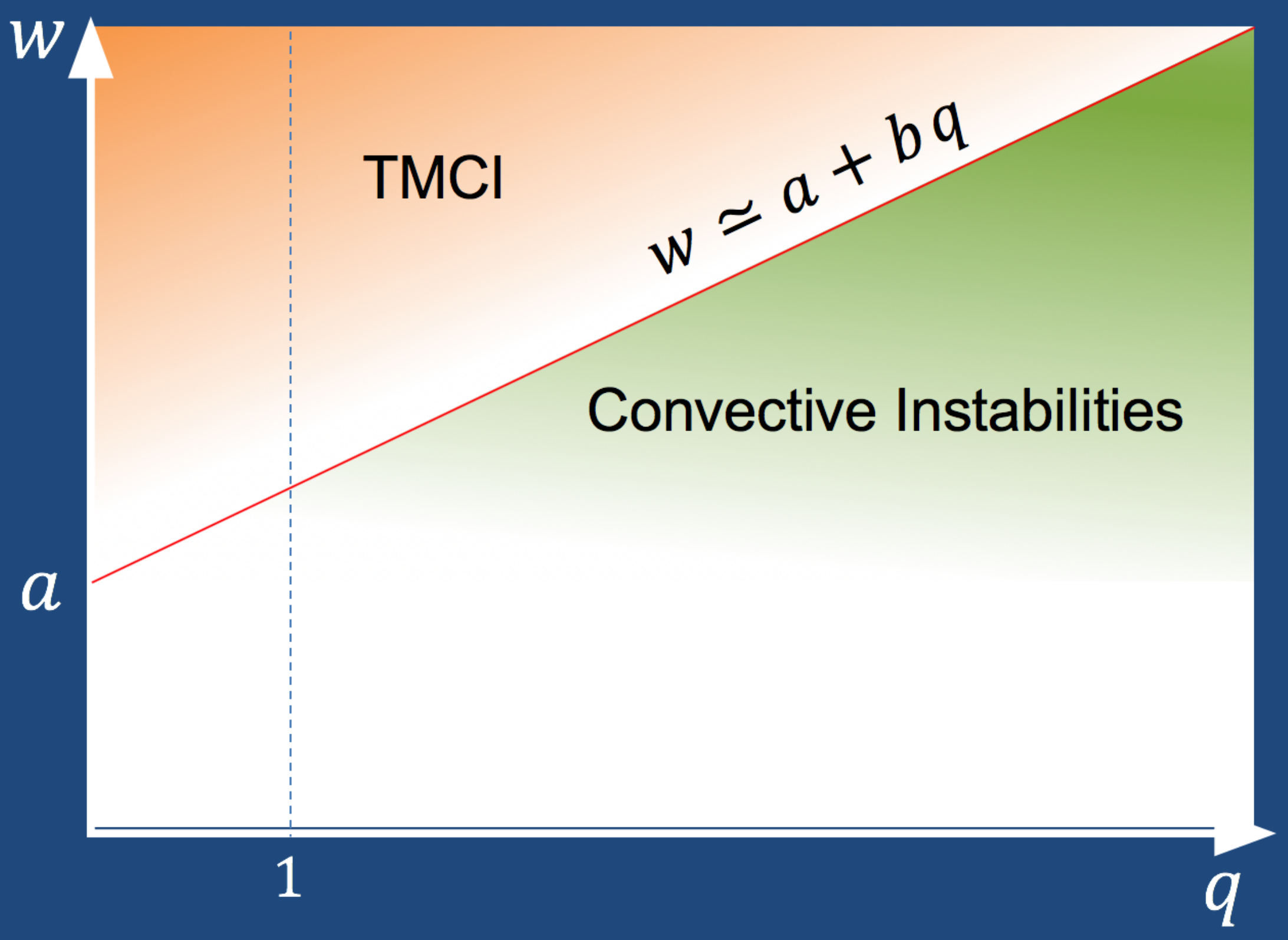}
\caption{\label{Fig:ParSpace}
	General sketch of the instability areas on the SC-wake plane. The color gradients represent the growth rate for the TMCI and the amplification for the SCI. The white area at $w \leq a$ corresponds to stability with insignificant amplification. The convective amplification grows exponentially with the wake amplitude; the TMCI threshold $w \simeq a+b\,q$ does not prevent the convective amplification to grow this way in the TMCI area as well. The coefficients $a=w_{th}^0$ and $b$ depend on the shape of the wake function.    
	}
\end{figure}
%


This presentation of the amplification indicates a slow convergence of the Fourier series for large amplification when the Fourier coefficients almost cancel each other in the denominator of Eq.~(\ref{LogK}). Thus, sufficiently many Fourier harmonics have to be kept to make the result correct; the higher amplification, the larger has to be the Fourier truncation number. By the same reason, the matrix elements $U_{lm}$ of Eq.~(\ref{Aeq}) have to be computed with extra accuracy, do not spoil such cancellations. These requirements are well-facilitated for the resonator wakes, since the integrals $U_{lm}$ can be taken analytically, allowing broad parameter scans to be reliable and reasonably fast even for huge amplifications. 

Figures~\ref{ContourKRateq0}--\ref{ContourKRateq10} present pairs of contour plots, for natural logarithm of the amplification, $\log K$, on the left, and for the growth rate $\Im \nu$, on the right, versus the wake parameter $0 < w< 100$, along the vertical axis, and the phase advance $0<k_r<25$, along the horizontal. The last figure of this series, Fig.~\ref{ContourKq20}, shows only the amplification, since there is no TMCI there, $\Im \nu =0$ for the entire area of parameters at its strongest SC, $q=20$. The black dashed line on some of the plots shows the ABS no-SC TMCI threshold, Eq.~(\ref{NoSCTMCI}), in agreement with the TMCI border of Fig.~\ref{ContourKRateq0} for the phase advances $k_r \geq 3$. The left part of this figure shows that amplification can be large in the absolutely-stable area even without SC; the higher phase advance $k_r$, the larger can be the amplification of the absolutely-stable bunch. Thus, in principle, for very short wakes the convective instabilities may be more dangerous than absolute, even without SC. All these contour plots illustrate how the absolute instability, TMCI, recedes with growing SC, and the amplification increases along its no-SC threshold line. For the strongest SC case, $q=20$ of Fig.~\ref{ContourKq20}, modeling CERN SPS Q26 situation, the no-SC threshold line almost goes along the amplification level line $K \simeq 300\, - 1000$ for as short wakes as supposed to be at the SPS. Observations, presented in Ref.~\cite{Bartosik:2013qji}, Fig.~4.21 top left therein, seem to be compatible with this estimation. This explains why the mistaken assumption of TMCI insensitivity on SC worked fairly well for prediction of the intensity limitations at the SPS. At Q26, the machine was limited not by TMCI, which threshold was far above, but by amplification of the convective instability, which physically acceptable limit of $\simeq 1000$ occured fairly close to the no-SC threshold. For Q20 optics with its moderate SC parameter $q \approx 5$, the no-SC TMCI threshold was not that far, $\simeq 20-30\%$ below its actual threshold, as one may see in Fig.~\ref{ContourKRateq5} with $k_r \simeq 20-25$.   

Figure~\ref{Fig:ParSpace} suggests a general schematic plot for the TMCI and SCI areas on the SC-wake plane, where the color intensity varies either with the growth rate (orange) or the amplification (green). According to Ref.~\cite{PhysRevAccelBeams.21.104201}, at sufficiently large SC parameter, the TMCI threshold $w_\mathrm{th}$ increases linearly with that, $w_\mathrm{th} \propto q$, for all practically important cases. That is why the threshold is represented by a straight line at the sketch. The reader should not be confused by the green area border from above: the TMCI threshold does not prevent the convective amplification to grow exponentially with the wake in the TMCI area as well. For the broadband wake cases, demonstrated in Figs.~\ref{ContourKRateq0}-\ref{ContourKRateq10}, the threshold slope is estimated as $b \simeq 4$ for all $k_r \geq 6$, while its no-SC value $a=w_\mathrm{th}^0$ is given by Eq.~\ref{NoSCTMCI}.

\section{\label{sec:OS} Theory, Observations and Simulations}

Many impressive observations of the convective instabilities were actually made at the CERN machines, albeit the instabilities were usually misinterpreted as TMCI. For instance, Fig.~5 of Ref.~\cite{Quatraro:2010hza} shows a convective signal at the PSB, with head-to-tail amplification not less than $\sim 10$, as it may be guessed at a glance. Much larger amplification was observed at the PS, see e.g. Fig.~14 of Ref.~\cite{Metral:2016grn}, where the amplification looks like it is in the range hundreds, if not more. A huge head-to-tail amplification is seen in Fig.~4.21, top left, of Ref.~\cite{Bartosik:2013qji}, showing bunch transverse oscillations measured at the SPS. To the right of this figure, a result of the no-SC HEADTAIL simulations is presented as a counterpart, with a claim that the two plots show "very similar intra bunch motion". It is hard to agree with this claim though, seeing an enormous measured amplification on the left and almost perfectly mirror-symmetric simulation picture on the right. It deserves a special reflection, that although no model was ever suggested with a considerable amplification generated by the mode coupling, and no cases were theoretically found with mode coupling insensitive to SC, still the single bunch instabilities at strong SC, with their impressive head-to-tail amplification, were unanimously called "TMCI" in countless publications, and in many of them the no-SC formula for the instability threshold, never derived for strong SC, was treated as a theoretical "result" for cases with strong SC.

To further complete the picture, more about its theoretical aspect has to be said. As it was already discussed in the beginning of this paper, one of the firmly established theoretical conclusions about TMCI was independence of the stability condition on the number of particles at the strong SC case: if the bunch is stable at some population, it must be stable at a higher intensity, regardless of how much higher! Apparently, this amazing statement was never publicly criticized or rejected as obviously unreasonable. When there was a need to compare the observations, like those mentioned above, with theoretical predictions, the only theory available so far was no-SC TMCI model in its various implementations, not much different from each other. The no-SC TMCI theory was applied to beams with large SC parameters not only without any theoretical justification, but against theoretical conclusions, which unanimously~\cite{blaskiewicz1998fast, burov2009head, PhysRevSTAB.14.094401} cried out about the opposite, that strong SC does change the situation dramatically. The theoretically grounded conclusions were tacitly disregarded, and the fully ungrounded statement of validity of no-SC theory for strong SC cases was employed for checking this sort of observations with the "theory" or with whatever stood for it. This comparison had one serious justification, mentioned in the previous section: the model at hand did predict intensity limitations fairly well. For such a virtue, many theoretical sins can be excused by practical people, especially when their only choice is between an unjustified, but at least partly working formula and a failure, if not absurdity, even if the latter was apparently derived from the first principles. This checking with theory was also complemented by a remarkable agreement between different no-SC computations, like MOSES and HEADTAIL~\cite{Salvant:2010dda, Metral:2016grn}, convincing that both no-SC programs are most likely correct, not more.  Well, turning a blind eye to theoretical inconsistency can be comprehensible in this kind of situation, but it may cost progress in understanding, since the value of the latter is supplanted by too empiricist attitude. 

A seed of new understanding can be seen in a publication of D.~Quatraro and G.~Rumolo~\cite{quatraro2010effects}, where significant dependence of the "TMCI" threshold on SC was demonstrated, see Fig.~4 therein. Figure~3 of that article, computed with the resistive wall impedance, clearly shows that all the excited modes are positive, contrary to the no-SC situation. It shows also that the modes are uncoupled at the "mode coupling threshold." Moreover, it shows that the mode excitation gradually increases with intensity, rather than suddenly springing from a barely visible state of stability to the infinite radiance of the real TMCI with no SC, as in Fig.~5 therein. All these important features of the collective dynamics, clearly seen in the pioneer results of Ref.~\cite{quatraro2010effects}, create the impression that only a tiny step separated its authors from overcoming the common misconceptions and discovering the new types of beam collective instabilities.

\begin{acknowledgments}

I am grateful to Elias Metral for the long-term pleasure of enlightening and provocative discussions, in general, and for pointing my attention to the publication of D.~Quatraro and G.~Rumolo~\cite{quatraro2010effects}, in particular. I am thankful to one of the Referees of this article, who pointed my attention to the similar problems discussed for the longitudinal plane. I am grateful to Tim Zolkin for his help in early stages of this work. My special thanks are to Slava Derbenev for his inspiring interest in this research and its high evaluation. 

Fermilab is operated by Fermi Research Alliance, LLC under Contract No. DE-AC02-07CH11359 with the United States Department of Energy.

\end{acknowledgments}

\bibliography{bibfile}			
\end{document}